\documentclass[
aps,
prd,
preprint,
superscriptaddress,
nofootinbib,
longbibliography
]{revtex4-2}

\usepackage{amsmath}
\usepackage{amssymb}
\usepackage{bm}
\usepackage{mathtools}
\usepackage{graphicx}
\usepackage{xcolor}
\usepackage{microtype}
\usepackage[hidelinks]{hyperref}

\hypersetup{
  pdftitle={
    Exact Phenomenology of the Neutrino Cuboid:
    Normal Mass Ordering, the Tribimaximal Limit,
    and Cosmological Constraints
  },
  pdfauthor={Jianlong Lu},
  pdfsubject={
    Exact mass--mixing relations and phenomenology
    of the neutrino cuboid
  },
  pdfkeywords={
    neutrino masses,
    neutrino mixing,
    neutrino mass ordering,
    tribimaximal mixing,
    cosmological neutrino mass
  }
}

\begin{document}

\title{Exact Phenomenology of the Neutrino Cuboid: Normal Mass Ordering, the Tribimaximal Limit, and Cosmological Constraints}

\author{Jianlong Lu}
\email{jianlong@nus.edu.sg}
\affiliation{
Department of Mathematics, National University of Singapore,
Singapore 119076
}

\date{\today}

\begin{abstract}
We investigate the exact phenomenology of a geometric neutrino cuboid defined by $m_1=m_0\sin\xi$, $m_2=m_0\cos\xi\sin\zeta$, and $m_3=m_0\cos\xi\cos\zeta$, whose cubic point corresponds to mass degeneracy and the tribimaximal values of the solar and atmospheric mixing angles. Imposing the mass–mixing alignment $\xi=\theta_{12}$ and $\zeta=\theta_{23}$, we derive closed-form consistency relations without relying on an expansion about the cubic limit. These relations show that the observed condition $\sin^2\theta_{12}<1/3$ selects normal mass ordering and, in the resulting normal-ordering regime, the physical condition $0<\Delta m_{21}^2/\Delta m_{31}^2<1$ requires $\theta_{23}$ to lie in the lower octant. Representative JUNO-based inputs give $\sin^2\theta_{23}\simeq0.440$, $(m_1,m_2,m_3)\simeq(0.0938,0.0942,0.1063)\,\mathrm{eV}$, and $\sum_i m_i\simeq0.294\,\mathrm{eV}$. We demonstrate that the first-order expansion around the tribimaximal point is numerically unstable for the solar-to-atmospheric mass-splitting ratio because of a leading-order cancellation. Present atmospheric-angle data yield only mild tension with exact alignment, whereas stringent neutrino-mass bounds in baseline $\Lambda$CDM cosmology strongly disfavor it; its viability under extended cosmological models remains model dependent. We finally formulate geometric alignment residuals as general null-test observables, providing a systematic framework for testing the neutrino cuboid with future oscillation and absolute-mass measurements.
\end{abstract}


\maketitle

\clearpage

\section{Introduction}
\label{sec:introduction}

The observation of neutrino oscillations established that at least two neutrino mass eigenstates are nonzero and that lepton flavors mix, providing compelling evidence for physics beyond the minimal Standard Model \cite{SuperK1998,SNO2002}. Within the standard three-neutrino framework, oscillation phenomena are described by three mixing angles, one Dirac CP-violating phase, and two independent mass-squared differences. The discovery of a nonzero reactor angle $\theta_{13}$ completed the measurement of the three mixing angles \cite{DayaBay2012}, while increasingly precise global analyses have constrained most oscillation parameters at the few-percent level or better \cite{NuFIT2024,Capozzi2025}. Nevertheless, the origin of the observed flavor structure remains unknown. In particular, the neutrino mass ordering, the octant of the atmospheric mixing angle $\theta_{23}$, the absolute mass scale, the Majorana or Dirac nature of neutrinos, and the possible Majorana phases remain central open questions.

Medium-baseline reactor experiments provide an especially clean probe of the solar parameters and the neutrino mass ordering through the interference between oscillations governed by the solar and atmospheric mass-squared differences \cite{Zhan2008,Zhan2009,Li2013JUNO,JUNO2016}. Using its first $59.1$ days of data, the Jiangmen Underground Neutrino Observatory (JUNO) reported $\sin^2\theta_{12}=0.3092\pm0.0087$ and $\Delta m_{21}^2=(7.50\pm0.12)\times10^{-5}\,\mathrm{eV}^2$ under the normal-ordering hypothesis, already improving the precision of the corresponding pre-JUNO combination \cite{JUNO2026}. A subsequent preliminary larger-exposure update presented at Neutrino 2026 further illustrates the rapid approach toward precision tests of the three-neutrino framework \cite{WangNeutrino2026}. Together with reactor determinations of the atmospheric mass-splitting scale and accelerator and atmospheric measurements of $\theta_{23}$, these results sharpen the experimentally accessible ratio $\eta\equiv\Delta m_{21}^2/\Delta m_{31}^2$, whose small value, $\eta\simeq3\times10^{-2}$ for normal ordering, provides a sensitive discriminator among proposed relations between neutrino masses and mixing angles.

A historically important organizing principle for lepton flavor mixing is the tribimaximal pattern \cite{Harrison2002,HeZee2003}. It predicts $\sin^2\theta_{12}=1/3$, $\sin^2\theta_{23}=1/2$, and $\theta_{13}=0$, and it can arise from a variety of discrete flavor symmetries and residual-symmetry constructions, including models based on $A_4$, $S_4$, and approximate $\mu$–$\tau$ symmetry \cite{BabuMaValle2003,XingZhao2016,Lu2025ICHEP,Lu2024Universe}. The measured nonzero value of $\theta_{13}$ excludes exact tribimaximal mixing as a complete description of the lepton mixing matrix, but the tribimaximal values of the two large angles remain useful zeroth-order reference points for organizing symmetry-breaking effects. Likewise, nearly degenerate neutrino masses can be compatible with permutation or other non-Abelian flavor symmetries \cite{FritzschXing1996,Jora2006}. Exact mass degeneracy alone, however, does not determine the oscillation mixing angles, since oscillations disappear when all mass-squared differences vanish. A predictive connection between mass degeneracy and a particular mixing pattern therefore requires an additional alignment condition or a specified symmetry-breaking structure.

Flavor models frequently produce testable sum rules. Conventional neutrino mass sum rules relate the generally complex Majorana mass eigenvalues and thereby constrain the Majorana phases and the effective mass relevant to neutrinoless double-beta decay \cite{Spinrath2016}. Mixing sum rules instead correlate observable mixing angles and the Dirac phase, often by treating tribimaximal or another symmetric pattern as the neutrino-sector limit and including charged-lepton corrections \cite{Antusch2007}. Relations involving both masses and mixing parameters occur less frequently and are normally tied to a particular dynamical framework, such as a grand-unified or flavor-symmetry construction \cite{Buccella2017}. Their phenomenological usefulness depends crucially on whether the proposed relation is treated exactly or through a controlled approximation, especially when a small observable results from cancellation between individually larger contributions.

A recent proposal introduced a geometric ``neutrino cuboid'' in which the three positive neutrino masses are parametrized as $m_1=m_0\sin\xi$, $m_2=m_0\cos\xi\sin\zeta$, and $m_3=m_0\cos\xi\cos\zeta$ \cite{Xing2026Cuboid}. These relations imply $m_0^2=m_1^2+m_2^2+m_3^2$ and amount to spherical coordinates for the normalized mass vector in its positive octant. At the cubic point $\xi_*=\arctan(1/\sqrt{2})$ and $\zeta_*=\pi/4$, one obtains $m_1=m_2=m_3=m_0/\sqrt{3}$. The same two geometric angles coincide with the solar and atmospheric values of the tribimaximal pattern, namely $\theta_{12}=\arctan(1/\sqrt{2})$ and $\theta_{23}=\pi/4$. This correspondence concerns the two large mixing angles; the reactor angle $\theta_{13}$ is not fixed by the cuboid construction itself. Moreover, the geometric parametrization alone is a change of variables and carries no physical prediction until a relation between the geometric and oscillation angles is imposed.

The central ansatz of Ref.~\cite{Xing2026Cuboid} is the exact mass–mixing alignment $\xi=\theta_{12}$ and $\zeta=\theta_{23}$. Expanding about the cubic point then suggests a connection among the departures from the tribimaximal values and the ratio $\eta$. This proposal is particularly interesting because two measured mixing angles would determine not only the shape of the mass spectrum but also its absolute scale. At the same time, the small solar mass splitting arises from a cancellation between the deviations associated with $\xi$ and $\zeta$. Consequently, a nominally first-order expansion of the individual masses need not yield a first-order-accurate prediction for $\eta$: terms that are formally quadratic in the angular deviations can become numerically comparable to the residual first-order numerator. An exact treatment is therefore required before the ansatz can be meaningfully confronted with precision data.

The absolute mass scale provides an independent and highly complementary test. Direct tritium beta-decay spectroscopy currently gives the model-independent bound $m_\beta<0.45\,\mathrm{eV}$ at $90\%$ C.L.~\cite{KATRIN2025}. If neutrinos are Majorana particles, neutrinoless double-beta decay constrains the coherent effective mass $m_{\beta\beta}$, subject to the Majorana phases, the assumed decay mechanism, and nuclear-matrix-element uncertainties \cite{KamLANDZen2025}. Cosmological observations provide the strongest current limits on the mass sum $\Sigma_\nu\equiv m_1+m_2+m_3$, but these limits depend on the cosmological model, datasets, priors, and treatment of systematic uncertainties \cite{Planck2018Cosmology,DESI2025Neutrino,Jimenez2026}. The different theoretical assumptions entering these observables make their combined comparison particularly valuable.

In this work, we develop the exact phenomenology of the neutrino-cuboid alignment and extend it into a general geometric null-test framework. First, we solve the alignment conditions analytically, without expanding about the cubic point. In particular, defining $x\equiv\sin^2\theta_{12}$, $y\equiv\sin^2\theta_{23}$, and $\eta\equiv\Delta m_{21}^2/\Delta m_{31}^2$, we obtain $m_0^2=(\Delta m_{21}^2+\Delta m_{31}^2)/(1-3x)$ and an exact relation between $x$, $y$, and $\eta$. Positivity of $m_0^2$, together with the observed condition $x<1/3$, selects normal mass ordering within the exact alignment hypothesis. We further derive the identity $y-1/2=(3x-1)(1-\eta)/[2(1-x)(1+\eta)]$, which shows that $x<1/3$ and $0<\eta<1$ necessarily place $\theta_{23}$ in the lower octant. Thus normal ordering, the atmospheric octant, and the absolute mass scale are correlated consequences of a single alignment assumption rather than independent inputs.

Second, we compare the exact relations with their first- and second-order expansions around the cubic point and identify the cancellation that makes the first-order prediction for the solar-to-atmospheric splitting ratio numerically unstable. Representative JUNO-based inputs lead to a normal and nearly degenerate spectrum with $(m_1,m_2,m_3)\simeq(0.0938,0.0942,0.1063)\,\mathrm{eV}$ and $\Sigma_\nu\simeq0.294\,\mathrm{eV}$. The corresponding prediction $\sin^2\theta_{23}\simeq0.440$ is only mildly disfavored by present oscillation data, whereas the predicted mass sum is strongly incompatible with the most restrictive baseline-$\Lambda$CDM bounds. This distinction is important: current oscillation measurements do not by themselves exclude the exact alignment, while its cosmological viability is highly conditional on the assumed cosmological framework.

Finally, we define the geometric alignment residuals $R_\xi\equiv\arcsin(m_1/m_0)-\theta_{12}$ and $R_\zeta\equiv\arctan(m_2/m_3)-\theta_{23}$, where $m_0=\sqrt{m_1^2+m_2^2+m_3^2}$. The exact cuboid ansatz corresponds to the point $(R_\xi,R_\zeta)=(0,0)$, while oscillation and absolute-mass measurements constrain the surrounding residual plane. This formulation separates the kinematic geometry from the dynamical hypothesis of exact alignment and permits controlled tests of approximate alignment, experimental uncertainties, and possible symmetry-breaking effects.

The remainder of this paper is organized as follows. In Sec.~\ref{sec:geometry}, we formulate the neutrino-cuboid geometry and derive the exact mass–mixing sum rules. In Sec.~\ref{sec:expansion}, we analyze the expansion about the cubic and tribimaximal limit and quantify the cancellation affecting the solar mass splitting. In Sec.~\ref{sec:oscillation}, we confront the exact alignment and its residual generalization with oscillation data. Section~\ref{sec:absolute_mass} discusses the predictions for $m_\beta$, $m_{\beta\beta}$, and $\Sigma_\nu$, together with laboratory and cosmological constraints. We summarize our main conclusions and future tests in Sec.~\ref{sec:conclusions}.

\section{Neutrino-cuboid geometry and exact mass–mixing sum rules}
\label{sec:geometry}

\subsection{Geometric parametrization}

Let $m_i>0$, with $i=1,2,3$, denote the three physical neutrino masses. The neutrino-cuboid parametrization introduced in Ref.~\cite{Xing2026Cuboid} is
\begin{equation}
  m_1=m_0\sin\xi,
  \qquad
  m_2=m_0\cos\xi\sin\zeta,
  \qquad
  m_3=m_0\cos\xi\cos\zeta,
  \label{eq:cuboid-parametrization}
\end{equation}
where $m_0>0$ and $0<\xi,\zeta<\pi/2$. The strict inequalities $m_i>0$ place the geometric angles in the interior of this angular domain. Boundary spectra with one vanishing mass, used later in constructing the residual trajectories, are understood as continuous limits of the positive-mass domain. Squaring and adding the three relations gives
\begin{equation}
m_0^2=m_1^2+m_2^2+m_3^2.
\label{eq:m0-norm}
\end{equation}
Thus, $m_0$ is the Euclidean norm of the neutrino mass vector and should not be confused with the lightest neutrino mass. For positive masses, Eq.~\eqref{eq:cuboid-parametrization} can be inverted uniquely:
\begin{equation}
  \begin{aligned}
    \xi
    &=
    \arcsin\left(\frac{m_1}{m_0}\right)
    =
    \arctan\left(\frac{m_1}{\sqrt{m_2^2+m_3^2}}\right),
    \\
    \zeta
    &=
    \arctan\left(\frac{m_2}{m_3}\right).
  \end{aligned}
  \label{eq:inverse-geometric-angles}
\end{equation}
Consequently, Eq.~\eqref{eq:cuboid-parametrization} is, by itself, an exact reparametrization of three positive masses and does not reduce the number of independent mass parameters. Its predictive content arises only after a relation between the geometric angles $\xi$ and $\zeta$ and independently measurable flavor-mixing parameters is imposed.

The cubic point is defined by $m_1=m_2=m_3$. Equations~\eqref{eq:cuboid-parametrization} and \eqref{eq:m0-norm} then imply
\begin{equation}
  \begin{aligned}
    m_1 &= m_2=m_3=\frac{m_0}{\sqrt{3}},
    \\
    \xi_*
    &=
    \arcsin\left(\frac{1}{\sqrt{3}}\right)
    =
    \arctan\left(\frac{1}{\sqrt{2}}\right),
    \qquad
    \zeta_*=\frac{\pi}{4}.
  \end{aligned}
  \label{eq:cubic-point}
\end{equation}
The values $\xi_*$ and $\zeta_*$ coincide with the solar and atmospheric angles of tribimaximal mixing, for which $\sin^2\theta_{12}=1/3$ and $\sin^2\theta_{23}=1/2$. This correspondence does not determine the reactor angle $\theta_{13}$, so the cuboid geometry alone does not reproduce the complete tribimaximal mixing matrix. Moreover, exact mass degeneracy does not by itself select any observable oscillation mixing pattern, since all vacuum oscillation phases vanish when the mass-squared differences are zero. A physical connection between the cubic mass limit and tribimaximal flavor mixing therefore requires an additional alignment hypothesis.

\subsection{Exact mass–mixing alignment}

We impose the alignment proposed in Ref.~\cite{Xing2026Cuboid},
\begin{equation}
  \xi=\theta_{12},
  \qquad
  \zeta=\theta_{23}.
  \label{eq:exact-alignment}
\end{equation}
For compactness, we define
\begin{equation}
  x\equiv\sin^2\theta_{12},
  \qquad
  y\equiv\sin^2\theta_{23}.
  \label{eq:x-y-definitions}
\end{equation}

The squared masses following from Eqs.~\eqref{eq:cuboid-parametrization} and \eqref{eq:exact-alignment} are then
\begin{equation}
  m_1^2=m_0^2x,
  \qquad
  m_2^2=m_0^2(1-x)y,
  \qquad
  m_3^2=m_0^2(1-x)(1-y).
  \label{eq:aligned-masses-squared}
\end{equation}

We adopt the signed mass-squared differences
\begin{equation}
  \Delta m_{21}^2\equiv m_2^2-m_1^2>0,
  \qquad
  \Delta m_{31}^2\equiv m_3^2-m_1^2,
  \label{eq:mass-squared-differences}
\end{equation}
so that $\Delta m_{31}^2>0$ for normal ordering and $\Delta m_{31}^2<0$ for inverted ordering. Substitution of Eq.~\eqref{eq:aligned-masses-squared} gives
\begin{align}
  \Delta m_{21}^2
  &=
  m_0^2\left[(1-x)y-x\right],
  \label{eq:exact-dm21}
  \\
  \Delta m_{31}^2
  &=
  m_0^2\left[(1-x)(1-y)-x\right].
  \label{eq:exact-dm31}
\end{align}

Adding Eqs.~\eqref{eq:exact-dm21} and \eqref{eq:exact-dm31} eliminates $y$ and yields
\begin{equation}
  \Delta m_{21}^2+\Delta m_{31}^2
  =
  m_0^2(1-3x).
  \label{eq:sum-splittings}
\end{equation}

Provided $x\neq 1/3$, the norm of the mass vector is therefore fixed exactly:
\begin{equation}
  m_0^2
  =
  \frac{\Delta m_{21}^2+\Delta m_{31}^2}
       {1-3\sin^2\theta_{12}}.
  \label{eq:exact-m0}
\end{equation}

The three physical masses are consequently
\begin{align}
  m_1^2
  &=
  \frac{\sin^2\theta_{12}}
       {1-3\sin^2\theta_{12}}
  \left(\Delta m_{21}^2+\Delta m_{31}^2\right),
  \label{eq:exact-m1}
  \\
  m_2^2
  &=
  m_1^2+\Delta m_{21}^2,
  \label{eq:exact-m2}
  \\
  m_3^2
  &=
  m_1^2+\Delta m_{31}^2.
  \label{eq:exact-m3}
\end{align}

Equations~\eqref{eq:exact-m0}--\eqref{eq:exact-m3} show that the alignment $\xi=\theta_{12}$ removes the usual freedom associated with the lightest neutrino mass. Once $\theta_{12}$ and the two mass-squared differences are specified, the complete mass spectrum is fixed.

The singular behavior of Eq.~\eqref{eq:exact-m0} near the tribimaximal value has a transparent interpretation. If $\sin^2\theta_{12}$ approaches $1/3$ from below while the measured mass-squared differences are held fixed, then $m_0$ becomes large and the spectrum becomes increasingly degenerate. Exactly at $\sin^2\theta_{12}=1/3$, Eq.~\eqref{eq:sum-splittings} requires $\Delta m_{21}^2+\Delta m_{31}^2=0$. At the full cubic point, where $y=1/2$ as well, both mass-squared differences vanish. The exact cubic limit is therefore compatible only with exact mass degeneracy; it cannot coexist with the observed nonzero splittings at finite $m_0$.

To obtain the atmospheric angle, we define
\begin{equation}
  \eta
  \equiv
  \frac{\Delta m_{21}^2}{\Delta m_{31}^2}.
  \label{eq:eta-definition}
\end{equation}

For normal ordering, $\eta$ is positive and much smaller than unity. Dividing Eq.~\eqref{eq:exact-dm21} by Eq.~\eqref{eq:exact-dm31} gives the exact mass--mixing sum rule
\begin{equation}
  \eta
  =
  \frac{(1-x)y-x}
       {(1-x)(1-y)-x}
  =
  \frac{\sin^2\theta_{23}-\tan^2\theta_{12}}
       {\cos^2\theta_{23}-\tan^2\theta_{12}}.
  \label{eq:exact-eta-relation}
\end{equation}

Solving for $y$ yields
\begin{equation}
  \sin^2\theta_{23}
  =
  \frac{
    \tan^2\theta_{12}
    +
    \eta\left(1-\tan^2\theta_{12}\right)
  }
  {1+\eta},
  \label{eq:exact-theta23-tan}
\end{equation}
or equivalently,
\begin{equation}
  y
  =
  \frac{x+\eta(1-2x)}
       {(1-x)(1+\eta)}.
  \label{eq:exact-y}
\end{equation}

Thus, $\theta_{23}$ is not an independent parameter of the exact ansatz. It is predicted by $\theta_{12}$ and the ratio of the measured mass-squared differences.

\subsection{Ordering and octant theorems}

Equation~\eqref{eq:exact-m0} immediately connects the solar angle to the mass ordering. Since $m_0^2$ must be positive,
\begin{equation}
  \operatorname{sgn}
  \left(
    \Delta m_{21}^2+\Delta m_{31}^2
  \right)
  =
  \operatorname{sgn}
  \left(
    1-3\sin^2\theta_{12}
  \right).
  \label{eq:ordering-sign-condition}
\end{equation}

Experimentally, $\sin^2\theta_{12}<1/3$, so the denominator of Eq.~\eqref{eq:exact-m0} is positive. The exact alignment therefore requires
\begin{equation}
  \Delta m_{21}^2+\Delta m_{31}^2>0.
  \label{eq:normal-ordering-condition}
\end{equation}

This condition is automatically satisfied for normal ordering. For inverted ordering, however, $\Delta m_{31}^2<0$ and $\lvert\Delta m_{31}^2\rvert\gg\Delta m_{21}^2$, making the left-hand side of Eq.~\eqref{eq:normal-ordering-condition} negative. Hence the observed condition $\sin^2\theta_{12}<1/3$ excludes inverted ordering within the exact alignment hypothesis. Normal ordering is therefore an exact consequence of the alignment and the measured solar angle, rather than an additional assumption.

The atmospheric-octant prediction can also be written in a particularly transparent form. Subtracting $1/2$ from Eq.~\eqref{eq:exact-y} gives
\begin{equation}
  \sin^2\theta_{23}-\frac{1}{2}
  =
  \frac{
    \left(3\sin^2\theta_{12}-1\right)(1-\eta)
  }
  {
    2\cos^2\theta_{12}(1+\eta)
  }.
  \label{eq:exact-octant-identity}
\end{equation}

For the physical normal-ordering regime, $\sin^2\theta_{12}<1/3$ and $0<\eta<1$. The numerator on the right-hand side of Eq.~\eqref{eq:exact-octant-identity} is therefore negative, while the denominator is positive. It follows that
\begin{equation}
  \sin^2\theta_{23}<\frac{1}{2},
  \qquad
  \theta_{23}<\frac{\pi}{4}.
  \label{eq:lower-octant-prediction}
\end{equation}

The exact neutrino-cuboid alignment thus predicts the lower octant. Equality would require either $\sin^2\theta_{12}=1/3$ or $\eta=1$, neither of which is compatible with the observed oscillation parameters.

The sensitivity of the atmospheric prediction to the two inputs follows directly from Eq.~\eqref{eq:exact-y}:
\begin{equation}
  \begin{aligned}
    \frac{\partial y}{\partial x}
    &=
    \frac{1-\eta}
         {(1+\eta)(1-x)^2},
    \\
    \frac{\partial y}{\partial\eta}
    &=
    \frac{1-3x}
         {(1-x)(1+\eta)^2}.
  \end{aligned}
  \label{eq:y-sensitivities}
\end{equation}

Both derivatives are positive for $x<1/3$ and $0<\eta<1$. The prediction for $\sin^2\theta_{23}$ therefore increases monotonically with both $\sin^2\theta_{12}$ and $\eta$. Since $\eta$ is already small and comparatively well constrained, the uncertainty in the predicted atmospheric angle is expected to be dominated primarily by the uncertainty in $\theta_{12}$.

\subsection{Representative exact solution}

For direct comparison with Ref.~\cite{Xing2026Cuboid}, we use the representative inputs quoted there,
\begin{equation}
  \begin{aligned}
    \sin^2\theta_{12}
    &=
    0.3036,
    \\
    \Delta m_{21}^2
    &=
    7.388\times10^{-5}\,\mathrm{eV}^2,
    \\
    \Delta m_{31}^2
    &=
    2.509\times10^{-3}\,\mathrm{eV}^2.
  \end{aligned}
  \label{eq:representative-inputs}
\end{equation}

These values give
\begin{equation}
  \eta=0.029446,
  \qquad
  \theta_{12}=33.4356^\circ.
  \label{eq:representative-eta-theta12}
\end{equation}

The exact atmospheric-angle sum rule in Eq.~\eqref{eq:exact-y} then predicts
\begin{equation}
  \sin^2\theta_{23}=0.439620,
  \qquad
  \theta_{23}=41.5320^\circ.
  \label{eq:representative-theta23}
\end{equation}

The corresponding geometric deviations from the cubic point are
\begin{equation}
  \xi_*-\theta_{12}=1.8288^\circ,
  \qquad
  \zeta_*-\theta_{23}=3.4680^\circ.
  \label{eq:representative-deviations}
\end{equation}

Equations~\eqref{eq:exact-m0}--\eqref{eq:exact-m3} give
\begin{equation}
  m_0=0.170165\,\mathrm{eV},
  \label{eq:representative-m0}
\end{equation}
and
\begin{equation}
  \begin{aligned}
    m_1&=0.093761\,\mathrm{eV},
    \\
    m_2&=0.094154\,\mathrm{eV},
    \\
    m_3&=0.106302\,\mathrm{eV}.
  \end{aligned}
  \label{eq:representative-masses}
\end{equation}

The mass sum is therefore
\begin{equation}
  \Sigma_\nu
  \equiv
  m_1+m_2+m_3
  =
  0.294216\,\mathrm{eV}.
  \label{eq:representative-mass-sum}
\end{equation}

The spectrum is normal but only mildly hierarchical. In particular, the common mass scale of the two lighter states is substantially larger than $\sqrt{\Delta m_{31}^2}$, while the relative splitting between $m_1$ and $m_2$ is very small. The value of $m_0$ agrees with the approximate estimate $m_0\simeq 0.17\,\mathrm{eV}$ obtained in Ref.~\cite{Xing2026Cuboid}; the more consequential effect of the exact treatment concerns the correlation between the two angular deviations and the small ratio $\eta$, which will be examined in Sec.~\ref{sec:expansion}.

\subsection{Alignment residuals as null-test observables}

The cuboid geometry can be tested without assuming exact alignment from the outset. For any positive mass spectrum, we define the geometric angles
\begin{equation}
  \begin{aligned}
    \xi_{\mathrm{g}}
    &\equiv
    \arcsin\left(
      \frac{m_1}{\sqrt{m_1^2+m_2^2+m_3^2}}
    \right),
    \\
    \zeta_{\mathrm{g}}
    &\equiv
    \arctan\left(\frac{m_2}{m_3}\right).
  \end{aligned}
  \label{eq:geometric-angle-observables}
\end{equation}

The alignment residuals are then
\begin{equation}
  R_\xi\equiv\xi_{\mathrm{g}}-\theta_{12},
  \qquad
  R_\zeta\equiv\zeta_{\mathrm{g}}-\theta_{23}.
  \label{eq:alignment-residuals}
\end{equation}

The exact ansatz corresponds to the origin,
\begin{equation}
  (R_\xi,R_\zeta)=(0,0).
  \label{eq:alignment-origin}
\end{equation}

For normal ordering, the complete spectrum may be parametrized by the lightest mass $m_{\mathrm{lightest}}=m_1$:
\begin{equation}
  \begin{aligned}
    m_2
    &=
    \sqrt{m_1^2+\Delta m_{21}^2},
    \\
    m_3
    &=
    \sqrt{m_1^2+\Delta m_{31}^2}.
  \end{aligned}
  \label{eq:NO-spectrum-lightest}
\end{equation}

For inverted ordering, $m_{\mathrm{lightest}}=m_3$ and
\begin{equation}
  \begin{aligned}
    m_1
    &=
    \sqrt{m_3^2-\Delta m_{31}^2},
    \\
    m_2
    &=
    \sqrt{m_3^2-\Delta m_{31}^2+\Delta m_{21}^2},
  \end{aligned}
  \label{eq:IO-spectrum-lightest}
\end{equation}
where $\Delta m_{31}^2<0$. Varying the lightest mass therefore traces a one-dimensional trajectory in the $(R_\xi,R_\zeta)$ plane for each ordering. Oscillation measurements constrain the mixing-angle coordinates of this trajectory, while beta-decay, neutrinoless-double-beta-decay, and cosmological information constrain the allowed lightest-mass interval. The exact alignment point is reached only if both residuals vanish simultaneously.

This residual formulation makes explicit the distinction between the kinematic cuboid parametrization and the physical mass--mixing hypothesis. It also permits approximate alignment to be quantified continuously, rather than treating the exact ansatz as a purely binary proposition. In the following sections, we use the exact relations derived above both to assess the convergence of the expansion around the cubic point and to construct phenomenological tests of the alignment hypothesis.

\section{Controlled expansion about the cubic point}
\label{sec:expansion}

\subsection{Angular deviations and mass expansions}

To compare the exact relations with the expansion adopted in Ref.~\cite{Xing2026Cuboid}, we define the deviations from the cubic point as
\begin{equation}
  \epsilon_\xi
  \equiv
  \xi_*-\theta_{12},
  \qquad
  \epsilon_\zeta
  \equiv
  \zeta_*-\theta_{23}.
  \label{eq:angular-deviations}
\end{equation}
For the representative solution obtained in Sec.~\ref{sec:geometry}, both deviations are positive. All angular deviations appearing in the following expansions are understood to be expressed in radians. We use $\epsilon\equiv\max\left(\lvert\epsilon_\xi\rvert,\lvert\epsilon_\zeta\rvert\right)$ to denote their common formal order.

Under the exact alignment, the geometric angles are
\begin{equation}
  \xi=\xi_*-\epsilon_\xi,
  \qquad
  \zeta=\zeta_*-\epsilon_\zeta.
  \label{eq:angles-near-cubic-point}
\end{equation}
Expanding the three masses in Eq.~\eqref{eq:cuboid-parametrization} through second order gives
\begin{equation}
  \frac{m_1}{m_0}
  =
  \frac{1}{\sqrt{3}}
  \left(
    1-\sqrt{2}\,\epsilon_\xi
    -\frac{1}{2}\epsilon_\xi^2
  \right)
  +
  \mathcal{O}(\epsilon^3),
  \label{eq:m1-second-order}
\end{equation}
\begin{align}
  \frac{m_2}{m_0}
  &=
  \frac{1}{\sqrt{3}}
  \left(
    1+\frac{\epsilon_\xi}{\sqrt{2}}
    -\epsilon_\zeta
    -\frac{1}{2}\epsilon_\xi^2
    -\frac{\epsilon_\xi\epsilon_\zeta}{\sqrt{2}}
    -\frac{1}{2}\epsilon_\zeta^2
  \right)
  +
  \mathcal{O}(\epsilon^3),
  \label{eq:m2-second-order}
\end{align}
and
\begin{align}
  \frac{m_3}{m_0}
  &=
  \frac{1}{\sqrt{3}}
  \left(
    1+\frac{\epsilon_\xi}{\sqrt{2}}
    +\epsilon_\zeta
    -\frac{1}{2}\epsilon_\xi^2
    +\frac{\epsilon_\xi\epsilon_\zeta}{\sqrt{2}}
    -\frac{1}{2}\epsilon_\zeta^2
  \right)
  +
  \mathcal{O}(\epsilon^3).
  \label{eq:m3-second-order}
\end{align}
Retaining only the terms linear in $\epsilon_\xi$ and $\epsilon_\zeta$ reproduces the first-order mass formulas of Ref.~\cite{Xing2026Cuboid}. Although the two deviations are numerically small, this fact alone does not guarantee that every observable constructed from the masses is accurately described at first order. In particular, an observable whose leading contribution is a difference between two terms of comparable magnitude can receive a parametrically enhanced relative correction.

\subsection{Expansion of the mass-squared differences}

For convenience, we introduce the dimensionless splittings
\begin{equation}
  d_{21}
  \equiv
  \frac{\Delta m_{21}^2}{m_0^2},
  \qquad
  d_{31}
  \equiv
  \frac{\Delta m_{31}^2}{m_0^2}.
  \label{eq:dimensionless-splittings}
\end{equation}
Expanding the exact expressions in Eqs.~\eqref{eq:exact-dm21} and \eqref{eq:exact-dm31} gives
\begin{align}
  d_{21}
  &=
  \sqrt{2}\,\epsilon_\xi
  -\frac{2}{3}\epsilon_\zeta
  -\frac{1}{2}\epsilon_\xi^2
  -\frac{2\sqrt{2}}{3}
  \epsilon_\xi\epsilon_\zeta
  +
  \mathcal{O}(\epsilon^3),
  \label{eq:dm21-second-order}
\end{align}
and
\begin{align}
  d_{31}
  &=
  \sqrt{2}\,\epsilon_\xi
  +\frac{2}{3}\epsilon_\zeta
  -\frac{1}{2}\epsilon_\xi^2
  +\frac{2\sqrt{2}}{3}
  \epsilon_\xi\epsilon_\zeta
  +
  \mathcal{O}(\epsilon^3).
  \label{eq:dm31-second-order}
\end{align}
No term proportional to $\epsilon_\zeta^2$ appears at this order. This follows from the expansions $\sin^2(\pi/4-\epsilon_\zeta)=1/2-\epsilon_\zeta+\mathcal{O}(\epsilon_\zeta^3)$ and $\cos^2(\pi/4-\epsilon_\zeta)=1/2+\epsilon_\zeta+\mathcal{O}(\epsilon_\zeta^3)$.

Adding and subtracting Eqs.~\eqref{eq:dm21-second-order} and \eqref{eq:dm31-second-order} gives
\begin{equation}
  \frac{
    \Delta m_{21}^2+\Delta m_{31}^2
  }{m_0^2}
  =
  2\sqrt{2}\,\epsilon_\xi
  -
  \epsilon_\xi^2
  +
  \mathcal{O}(\epsilon^3),
  \label{eq:sum-splittings-second-order}
\end{equation}
and
\begin{equation}
  \frac{
    \Delta m_{31}^2-\Delta m_{21}^2
  }{m_0^2}
  =
  \frac{4}{3}\epsilon_\zeta
  +
  \frac{4\sqrt{2}}{3}
  \epsilon_\xi\epsilon_\zeta
  +
  \mathcal{O}(\epsilon^3).
  \label{eq:difference-splittings-second-order}
\end{equation}
Equation~\eqref{eq:sum-splittings-second-order} is also obtained directly by expanding the exact identity $1-3\sin^2(\xi_*-\epsilon_\xi)$. The terms involving $\epsilon_\zeta$ cancel from the sum, explaining why the absolute mass scale can remain stable even when the solar splitting is not accurately reproduced at first order.

\subsection{Cancellation in the solar splitting}

At first order, Eqs.~\eqref{eq:dm21-second-order} and \eqref{eq:dm31-second-order} reduce to
\begin{equation}
  d_{21}^{[1]}
  =
  \sqrt{2}\,\epsilon_\xi
  -
  \frac{2}{3}\epsilon_\zeta,
  \qquad
  d_{31}^{[1]}
  =
  \sqrt{2}\,\epsilon_\xi
  +
  \frac{2}{3}\epsilon_\zeta.
  \label{eq:first-order-splittings}
\end{equation}
The resulting first-order approximation to the splitting ratio is
\begin{equation}
  \eta_{[1]}
  \equiv
  \frac{d_{21}^{[1]}}{d_{31}^{[1]}}
  =
  \frac{
    3\sqrt{2}\,\epsilon_\xi-2\epsilon_\zeta
  }{
    3\sqrt{2}\,\epsilon_\xi+2\epsilon_\zeta
  }.
  \label{eq:first-order-eta}
\end{equation}
The numerator becomes small near the cancellation direction
\begin{equation}
  \epsilon_\zeta
  =
  \frac{3}{\sqrt{2}}\epsilon_\xi.
  \label{eq:first-order-cancellation-line}
\end{equation}
The observed small value of $\eta$ places the physical solution close enough to this direction that the formally second-order terms in $d_{21}$ become important relative to its residual first-order value.

A necessary condition for the first-order solar splitting to be reliable is
\begin{equation}
  \left|
    \frac{1}{2}\epsilon_\xi^2
    +
    \frac{2\sqrt{2}}{3}
    \epsilon_\xi\epsilon_\zeta
  \right|
  \ll
  \left|
    \sqrt{2}\,\epsilon_\xi
    -
    \frac{2}{3}\epsilon_\zeta
  \right|.
  \label{eq:first-order-reliability-condition}
\end{equation}
This condition is substantially stronger than merely requiring $\lvert\epsilon_\xi\rvert\ll1$ and $\lvert\epsilon_\zeta\rvert\ll1$. It is not satisfied by the representative physical solution.

Keeping the splittings consistently through second order gives the improved ratio
\begin{align}
  \eta_{[2]}
  &\equiv
  \frac{d_{21}^{[2]}}{d_{31}^{[2]}}
  \notag\\
  &=
  \frac{
    \sqrt{2}\,\epsilon_\xi
    -\dfrac{2}{3}\epsilon_\zeta
    -\dfrac{1}{2}\epsilon_\xi^2
    -\dfrac{2\sqrt{2}}{3}
     \epsilon_\xi\epsilon_\zeta
  }{
    \sqrt{2}\,\epsilon_\xi
    +\dfrac{2}{3}\epsilon_\zeta
    -\dfrac{1}{2}\epsilon_\xi^2
    +\dfrac{2\sqrt{2}}{3}
     \epsilon_\xi\epsilon_\zeta
  }.
  \label{eq:second-order-eta}
\end{align}
Because both mass-squared differences vanish at the cubic point, $\eta$ takes the indeterminate form $0/0$ there and does not possess a direction-independent Taylor expansion about $(\epsilon_\xi,\epsilon_\zeta)=(0,0)$. Equation~\eqref{eq:second-order-eta} should therefore be understood as the ratio of the consistently truncated second-order splittings, rather than as an ordinary second-order Taylor series for $\eta$ itself.

\subsection{Numerical convergence}

For the representative exact solution in Eqs.~\eqref{eq:representative-eta-theta12}--\eqref{eq:representative-deviations}, the angular deviations are
\begin{equation}
  \begin{aligned}
    \epsilon_\xi
    &=
    0.0319187\,\mathrm{rad}
    =
    1.82881^\circ,
    \\
    \epsilon_\zeta
    &=
    0.0605276\,\mathrm{rad}
    =
    3.46798^\circ.
  \end{aligned}
  \label{eq:numerical-deviations-radians}
\end{equation}
At these values, the leading first-order terms in the solar splitting give
\begin{equation}
  \sqrt{2}\,\epsilon_\xi
  -
  \frac{2}{3}\epsilon_\zeta
  =
  4.78815\times10^{-3},
  \label{eq:numerical-leading-solar}
\end{equation}
whereas the combined second-order correction is
\begin{equation}
  -
  \frac{1}{2}\epsilon_\xi^2
  -
  \frac{2\sqrt{2}}{3}
  \epsilon_\xi\epsilon_\zeta
  =
  -2.33087\times10^{-3}.
  \label{eq:numerical-second-order-solar}
\end{equation}
The magnitude of the second-order correction is therefore approximately $49\%$ of the residual first-order contribution. By contrast, the corresponding second-order correction to $d_{31}$ is only about $1.5\%$ of its first-order value. The convergence problem is thus specific to the cancellation-suppressed solar splitting and is not a general failure of the small-angle expansion.

The exact and truncated results are compared in Table~\ref{tab:expansion-comparison}. For the mass scale, the first- and second-order approximations are defined by
\begin{equation}
  \left(m_0^{[1]}\right)^2
  =
  \frac{
    \Delta m_{21}^2+\Delta m_{31}^2
  }{
    2\sqrt{2}\,\epsilon_\xi
  },
  \label{eq:first-order-m0}
\end{equation}
and
\begin{equation}
  \left(m_0^{[2]}\right)^2
  =
  \frac{
    \Delta m_{21}^2+\Delta m_{31}^2
  }{
    2\sqrt{2}\,\epsilon_\xi-\epsilon_\xi^2
  }.
  \label{eq:second-order-m0}
\end{equation}

\begin{table}[t]
  \centering
  \caption{
    Comparison of the exact results with the first- and second-order truncations evaluated at the exact angular deviations in Eq.~\eqref{eq:numerical-deviations-radians}.
  }
  \label{tab:expansion-comparison}
  \small
  \setlength{\tabcolsep}{4pt}
  \begin{tabular}{lccc}
    \hline\hline
    Quantity
    &
    Exact
    &
    First order
    &
    Second order
    \\
    \hline
    $d_{21}$
    &
    $2.55145\times10^{-3}$
    &
    $4.78815\times10^{-3}$
    &
    $2.45727\times10^{-3}$
    \\
    $d_{31}$
    &
    $8.66485\times10^{-2}$
    &
    $8.54916\times10^{-2}$
    &
    $8.68037\times10^{-2}$
    \\
    $\eta$
    &
    $0.029446$
    &
    $0.056007$
    &
    $0.028308$
    \\
    $m_0/\mathrm{eV}$
    &
    $0.170165$
    &
    $0.169144$
    &
    $0.170107$
    \\
    \hline\hline
  \end{tabular}
\end{table}

The first-order approximation overestimates $d_{21}$ by approximately $88\%$ and predicts $\eta_{[1]}\simeq 0.0560$, almost twice its exact value. The second-order result reduces the relative error in $\eta$ to approximately $3.9\%$. In contrast, the first-order estimate of $m_0$ differs from the exact result by only about $0.6\%$, and the second-order estimate differs by less than $0.04\%$. This explains why the approximate value $m_0\simeq 0.17\,\mathrm{eV}$ remains accurate despite the instability of the first-order angular correlation.

\subsection{Inverse determination of the atmospheric deviation}

The sensitivity to truncation becomes even more pronounced when the first-order ratio is inverted to determine $\epsilon_\zeta$. Solving Eq.~\eqref{eq:first-order-eta} gives
\begin{equation}
  \epsilon_\zeta^{[1]}
  =
  \frac{3}{\sqrt{2}}
  \frac{1-\eta}{1+\eta}
  \epsilon_\xi.
  \label{eq:first-order-eps-zeta}
\end{equation}
For $\epsilon_\xi=0.0319187\,\mathrm{rad}$ and $\eta=0.029446$, this yields
\begin{equation}
  \epsilon_\zeta^{[1]}
  =
  0.0638363\,\mathrm{rad}
  =
  3.65755^\circ.
  \label{eq:first-order-eps-zeta-numerical}
\end{equation}
This differs from the exact value $\epsilon_\zeta=3.46798^\circ$ by only about $5.5\%$. Nevertheless, substituting $\epsilon_\zeta^{[1]}$ back into the exact relation in Eq.~\eqref{eq:exact-eta-relation} gives
\begin{equation}
  \eta_{\mathrm{exact}}
  \left(
    \epsilon_\xi,
    \epsilon_\zeta^{[1]}
  \right)
  =
  0.00298043,
  \label{eq:exact-eta-first-order-backsubstitution}
\end{equation}
which is almost one order of magnitude smaller than the input value $\eta=0.029446$. A modest angular error is thus amplified into a large relative error in the cancellation-suppressed quantity.

The second-order ratio can also be inverted analytically. Solving Eq.~\eqref{eq:second-order-eta} for $\epsilon_\zeta$ gives
\begin{equation}
  \epsilon_\zeta^{[2]}
  =
  \frac{3}{2}
  \frac{
    \sqrt{2}\,\epsilon_\xi
    -\epsilon_\xi^2/2
  }{
    1+\sqrt{2}\,\epsilon_\xi
  }
  \frac{1-\eta}{1+\eta}.
  \label{eq:second-order-eps-zeta}
\end{equation}
Numerically,
\begin{equation}
  \epsilon_\zeta^{[2]}
  =
  0.0603899\,\mathrm{rad}
  =
  3.46009^\circ,
  \label{eq:second-order-eps-zeta-numerical}
\end{equation}
which differs from the exact value by approximately $0.23\%$. Exact back-substitution using Eq.~\eqref{eq:exact-eta-relation} gives
\begin{equation}
  \eta_{\mathrm{exact}}
  \left(
    \epsilon_\xi,
    \epsilon_\zeta^{[2]}
  \right)
  =
  0.0305780,
  \label{eq:exact-eta-second-order-backsubstitution}
\end{equation}
showing that the second-order inversion captures the exact angular correlation much more reliably.

The exact cuboid alignment is therefore mathematically self-consistent, and its prediction for the absolute mass scale is stable. The limitation lies specifically in using the first-order expression for the cancellation-suppressed ratio $\eta$ as a precision relation. In the phenomenological analysis that follows, we consequently use the exact formulas of Sec.~\ref{sec:geometry}; the first- and second-order expressions are retained only to diagnose the structure and convergence of the expansion.

\section{Oscillation-data tests and alignment residuals}
\label{sec:oscillation}

\subsection{Exact prediction from oscillation inputs}

The exact cuboid alignment predicts the atmospheric angle from the solar angle and the ratio of mass-squared differences. Defining
\begin{equation}
F(x,\eta)
\equiv
\frac{x+\eta(1-2x)}
{(1-x)(1+\eta)},
\label{eq:alignment-map}
\end{equation}
the prediction in Eq.~\eqref{eq:exact-y} can be written as
\begin{equation}
y_{\mathrm{pred}}
\equiv
\sin^2\theta_{23}^{\mathrm{pred}}
=
F(x,\eta),
\qquad
\theta_{23}^{\mathrm{pred}}
=
\arcsin\!\sqrt{F(x,\eta)}.
\label{eq:predicted-theta23-map}
\end{equation}
Here, $x=\sin^2\theta_{12}$ and
$\eta=\Delta m_{21}^2/\Delta m_{31}^2$. The exact lower-octant
result in Eq.~\eqref{eq:lower-octant-prediction} follows immediately
from this map when $x<1/3$ and $0<\eta<1$.

It is useful to compare two solar-input benchmarks while using the
same atmospheric-scale input. The first is based on the published
analysis of the first $59.1$ days of JUNO data, which gives
\cite{JUNO2026}
\begin{equation}
\begin{aligned}
\mathcal{D}_{\mathrm{pub}}:\qquad
x
&=
0.3092\pm0.0087,
\\
\Delta m_{21}^2
&=
(7.50\pm0.12)
\times10^{-5}\,\mathrm{eV}^2.
\end{aligned}
\label{eq:published-JUNO-inputs}
\end{equation}
The label ``pub'' refers here only to the published solar inputs; the
atmospheric-scale input introduced below is taken from the later
preliminary JUNO--Daya Bay combination. For comparison, the
preliminary larger-exposure solar values quoted in
Ref.~\cite{Xing2026Cuboid} are \cite{WangNeutrino2026}
\begin{equation}
\begin{aligned}
\mathcal{D}_{\mathrm{pre}}:\qquad
x
&=
0.3036\pm0.0064,
\\
\Delta m_{21}^2
&=
(7.388\pm0.078)
\times10^{-5}\,\mathrm{eV}^2.
\end{aligned}
\label{eq:preliminary-JUNO-inputs}
\end{equation}
For both illustrative benchmarks, we use the atmospheric-scale input
quoted in the later preliminary JUNO analysis,
\begin{equation}
\Delta m_{31}^2
=
\left(
2.509^{+0.027}_{-0.025}
\right)
\times10^{-3}\,\mathrm{eV}^2,
\label{eq:atmospheric-splitting-input}
\end{equation}
obtained from a JUNO--Daya Bay combination
\cite{WangNeutrino2026,Xing2026Cuboid}. We symmetrize its uncertainty
to $\sigma_{31}=0.026\times10^{-3}\,\mathrm{eV}^2$ only for the
illustrative Gaussian propagation below.

The published-solar and preliminary inputs imply, respectively,
\begin{equation}
\begin{aligned}
\mathcal{D}_{\mathrm{pub}}:\qquad
\eta
&=
0.0298924,
&
y_{\mathrm{pred}}
&=
0.450639,
&
\theta_{23}^{\mathrm{pred}}
&=
42.1672^\circ,
\\
\mathcal{D}_{\mathrm{pre}}:\qquad
\eta
&=
0.0294460,
&
y_{\mathrm{pred}}
&=
0.439620,
&
\theta_{23}^{\mathrm{pred}}
&=
41.5320^\circ.
\end{aligned}
\label{eq:two-input-set-predictions}
\end{equation}
Both input sets therefore predict the lower octant, but the preliminary
shift of $x$ to a smaller value moves the predicted atmospheric angle
farther below maximal mixing.

\subsection{Propagation of oscillation uncertainties}

For a general covariance matrix, the uncertainty in the exact prediction
should be propagated through the nonlinear map in
Eq.~\eqref{eq:alignment-map}. At linear order,
\begin{equation}
\sigma_y^2
\simeq
\left(
\frac{\partial F}{\partial x}
\right)^2
\sigma_x^2
+
\left(
\frac{\partial F}{\partial\eta}
\right)^2
\sigma_\eta^2
+
2
\frac{\partial F}{\partial x}
\frac{\partial F}{\partial\eta}
\operatorname{Cov}(x,\eta),
\label{eq:y-error-propagation}
\end{equation}
where
\begin{equation}
\frac{\partial F}{\partial x}
=
\frac{1-\eta}
{(1+\eta)(1-x)^2},
\qquad
\frac{\partial F}{\partial\eta}
=
\frac{1-3x}
{(1-x)(1+\eta)^2}.
\label{eq:F-derivatives}
\end{equation}
If the two mass-squared differences are treated as independent, their
contribution to the uncertainty in $\eta$ is approximately
\begin{equation}
\sigma_\eta^2
\simeq
\eta^2
\left[
\left(
\frac{\sigma_{21}}{\Delta m_{21}^2}
\right)^2
+
\left(
\frac{\sigma_{31}}{\Delta m_{31}^2}
\right)^2
\right].
\label{eq:eta-error-propagation}
\end{equation}
The corresponding uncertainty in the angle, expressed in radians, is
\begin{equation}
\sigma_{\theta_{23}}
\simeq
\frac{\sigma_y}
{2\sqrt{y_{\mathrm{pred}}(1-y_{\mathrm{pred}})}}.
\label{eq:theta23-error-propagation}
\end{equation}

For the preliminary central values, the two derivatives are
\begin{equation}
\left.
\frac{\partial F}{\partial x}
\right|_{\mathrm{pre}}
=
1.9440,
\qquad
\left.
\frac{\partial F}{\partial\eta}
\right|_{\mathrm{pre}}
=
0.1209.
\label{eq:preliminary-F-derivatives}
\end{equation}
The atmospheric prediction is therefore much more sensitive to the solar
angle than to the already small and well-measured ratio $\eta$.

To account for the modest nonlinearity of $F(x,\eta)$, we also perform
an illustrative Monte Carlo propagation with $10^6$ independent
Gaussian samples for $x$, $\Delta m_{21}^2$, and the symmetrized
$\Delta m_{31}^2$. We use
\texttt{numpy.random.Generator} with the
\texttt{PCG64} generator and the fixed seed \texttt{20260724}.
The arrays for $x$, $\Delta m_{21}^2$, and $\Delta m_{31}^2$ are
generated in that order. Samples failing the physical-domain conditions
specified in Appendix~\ref{app:numerical} are discarded without
replacement, and empirical quantiles are evaluated using linear
interpolation. A complete covariance matrix covering all three inputs
in either hybrid benchmark, and in particular the parameters of the
preliminary update, is not publicly available. We therefore set the
correlations to zero only for this illustrative propagation. The
resulting medians and central $68\%$ intervals are summarized in
Table~\ref{tab:theta23-predictions}.

\begin{table}[t]
\centering
\caption{
Exact alignment predictions obtained from the published-solar and
preliminary JUNO-based input sets. The central values are empirical
medians, and the quoted intervals contain the central $68\%$ of the
illustrative Monte Carlo samples.
}
\label{tab:theta23-predictions}
\small
\setlength{\tabcolsep}{5pt}
\begin{tabular}{lcc}
\hline\hline
Input set
&
$\sin^2\theta_{23}^{\mathrm{pred}}$
&
$\theta_{23}^{\mathrm{pred}}$
\\
\hline
Published-solar
&
$0.4506^{+0.0172}_{-0.0168}$
&
$42.16^{+0.99}_{-0.97}{}^\circ$
\\
Preliminary
&
$0.4396^{+0.0125}_{-0.0122}$
&
$41.53^{+0.72}_{-0.71}{}^\circ$
\\
\hline\hline
\end{tabular}
\end{table}

For $\mathcal{D}_{\mathrm{pub}}$, the central $95\%$ intervals are
\begin{equation}
0.4177
<
\sin^2\theta_{23}^{\mathrm{pred}}
<
0.4843,
\qquad
40.27^\circ
<
\theta_{23}^{\mathrm{pred}}
<
44.10^\circ.
\label{eq:published-prediction-95}
\end{equation}
For $\mathcal{D}_{\mathrm{pre}}$, they are
\begin{equation}
0.4156
<
\sin^2\theta_{23}^{\mathrm{pred}}
<
0.4644,
\qquad
40.14^\circ
<
\theta_{23}^{\mathrm{pred}}
<
42.96^\circ.
\label{eq:preliminary-prediction-95}
\end{equation}
These intervals quantify only the propagation of the quoted
oscillation-input uncertainties. They do not include unknown
correlations, non-Gaussian experimental likelihoods, or systematic
differences between the published-solar benchmark and the preliminary
analysis.

\subsection{Comparison with global atmospheric-angle determinations}

Two representative global analyses give the normal-ordering results
\cite{Capozzi2025,NuFIT2024}
\begin{equation}
\begin{aligned}
\sin^2\theta_{23}
&=
0.473^{+0.023}_{-0.013},
&&
\text{Capozzi et al.},
\\
\sin^2\theta_{23}
&=
0.470^{+0.017}_{-0.013},
&&
\text{NuFIT}.
\end{aligned}
\label{eq:global-theta23-results}
\end{equation}
Both best-fit values lie in the lower octant, in qualitative agreement
with the exact octant theorem, but they are closer to maximal mixing
than the central cuboid predictions.

For orientation, we define an approximate Gaussian pull by
\begin{equation}
Z_{\mathrm{approx}}
\equiv
\frac{
y_{\mathrm{glob}}-y_{\mathrm{pred}}
}{
\sqrt{
\sigma_{y,\mathrm{pred},+}^2
+
\sigma_{y,\mathrm{glob},-}^2
}
},
\label{eq:approximate-theta23-pull}
\end{equation}
where $\sigma_{y,\mathrm{pred},+}$ is the upper-side uncertainty of the
alignment prediction and $\sigma_{y,\mathrm{glob},-}$ is the lower-side
uncertainty of the global result, because
$y_{\mathrm{pred}}<y_{\mathrm{glob}}$. This diagnostic gives
\begin{equation}
\begin{array}{c|cc}
&
\text{Capozzi et al.}
&
\text{NuFIT}
\\
\hline
\mathcal{D}_{\mathrm{pub}}
&
Z_{\mathrm{approx}}\simeq1.0
&
Z_{\mathrm{approx}}\simeq0.9
\\
\mathcal{D}_{\mathrm{pre}}
&
Z_{\mathrm{approx}}\simeq1.9
&
Z_{\mathrm{approx}}\simeq1.7
\end{array}.
\label{eq:approximate-pull-summary}
\end{equation}
The published-solar benchmark therefore exhibits no meaningful tension
with the global atmospheric-angle determinations. The preliminary input
set produces a somewhat deeper lower-octant prediction, but the
discrepancy remains at only the approximate $1.7\sigma$--$1.9\sigma$
level.

These numbers must not be interpreted as formal exclusion
significances. The likelihood for $\theta_{23}$ is generally asymmetric
and can contain octant-dependent structure; moreover, the oscillation
parameters entering the prediction and the global atmospheric fit need
not be statistically independent. The preliminary JUNO update also
does not provide the complete covariance and likelihood information
required for a rigorous combined analysis.

A proper likelihood test can be formulated by collecting the solar and
mass-splitting inputs into
$\mathbf{p}=(x,\Delta m_{21}^2,\Delta m_{31}^2)$ and profiling
\begin{equation}
\chi^2_{\mathrm{align}}
=
\min_{\mathbf{p}}
\left[
\chi^2_{\mathrm{input}}(\mathbf{p})
+
\Delta\chi^2_{23}
\left(
F\!\left(
x,
\frac{\Delta m_{21}^2}{\Delta m_{31}^2}
\right)
\right)
\right].
\label{eq:profiled-alignment-chi2}
\end{equation}
The goodness of fit of the alignment hypothesis should then be assessed
relative to the standard fit in which $\sin^2\theta_{23}$ is allowed to
vary independently. Evaluation of
Eq.~\eqref{eq:profiled-alignment-chi2} requires the experimental or
global-fit likelihood profiles and their relevant correlations, rather
than only the quoted one-dimensional intervals.

\subsection{Residual interpretation and future tests}

Conditioning on $R_\xi=0$ fixes the absolute mass scale through
Eq.~\eqref{eq:exact-m0} and predicts the geometric atmospheric angle.
The remaining residual is
\begin{equation}
R_\zeta
=
\theta_{23}^{\mathrm{pred}}
-
\theta_{23}^{\mathrm{glob}}.
\label{eq:conditional-R-zeta}
\end{equation}
Using the global best-fit values in
Eq.~\eqref{eq:global-theta23-results}, the central residuals are
\begin{equation}
\begin{aligned}
\mathcal{D}_{\mathrm{pub}}:\qquad
R_\zeta
&=
-1.29^\circ
\quad
\text{or}
\quad
-1.11^\circ,
\\
\mathcal{D}_{\mathrm{pre}}:\qquad
R_\zeta
&=
-1.92^\circ
\quad
\text{or}
\quad
-1.75^\circ,
\end{aligned}
\label{eq:central-R-zeta-values}
\end{equation}
where the first and second values correspond to the Capozzi et al.\
and NuFIT best fits, respectively.

The oscillation test of the exact alignment has a simple hierarchy.
First, a definitive determination of inverted ordering would exclude
the ansatz because Eq.~\eqref{eq:ordering-sign-condition} admits no
inverted-ordering solution for the observed value $x<1/3$. Second, a
definitive upper-octant result for $\theta_{23}$ would contradict the
exact identity in Eq.~\eqref{eq:exact-octant-identity}. Third, if normal
ordering and the lower octant are established, the continuous
correlation in Eq.~\eqref{eq:alignment-map} remains to be tested.

Near the preliminary central values, small changes obey
\begin{equation}
\delta y_{\mathrm{pred}}
\simeq
1.9440\,\delta x
+
0.1209\,\delta\eta.
\label{eq:future-linear-sensitivity}
\end{equation}
Thus, if the absolute uncertainty in $x$ reaches $\sigma_x=0.001$ while
the contribution from $\eta$ remains negligible, the alignment would
predict approximately
\begin{equation}
\sigma_y
\simeq
0.00194,
\qquad
\sigma_{\theta_{23}}
\simeq
0.11^\circ.
\label{eq:illustrative-future-precision}
\end{equation}
This illustrates the complementarity between increasingly precise
reactor measurements of $\theta_{12}$ and accelerator or atmospheric
determinations of $\theta_{23}$. The exact cuboid alignment is therefore
falsifiable using oscillation data alone, independently of the
cosmological tests considered in Sec.~\ref{sec:absolute_mass}.

\section{Absolute-mass observables and cosmological constraints}
\label{sec:absolute_mass}

A distinctive consequence of the exact alignment is that the absolute
neutrino-mass scale is no longer a free parameter. The oscillation
inputs determine not only the mass ordering and $\theta_{23}$ but also
the three individual masses, their sum, and the effective masses
relevant for beta decay and neutrinoless double-beta decay.

\subsection{Exact prediction for the neutrino-mass sum}

For compactness, we define
\begin{equation}
S
\equiv
\Delta m_{21}^2+\Delta m_{31}^2.
\label{eq:S-definition}
\end{equation}
Equations~\eqref{eq:exact-m1}--\eqref{eq:exact-m3} then give
\begin{equation}
m_1^2
=
\frac{xS}{1-3x},
\qquad
m_2^2
=
\frac{xS}{1-3x}
+
\Delta m_{21}^2,
\qquad
m_3^2
=
\frac{xS}{1-3x}
+
\Delta m_{31}^2.
\label{eq:exact-spectrum-compact}
\end{equation}
The exact mass sum is therefore
\begin{align}
\Sigma_\nu(x)
&\equiv
m_1+m_2+m_3
\notag\\
&=
\sqrt{
\frac{xS}{1-3x}
}
+
\sqrt{
\frac{xS}{1-3x}
+
\Delta m_{21}^2
}
\notag\\
&\quad
+
\sqrt{
\frac{xS}{1-3x}
+
\Delta m_{31}^2
}.
\label{eq:exact-neutrino-mass-sum}
\end{align}
This expression is valid throughout the physical normal-ordering
domain $0<x<1/3$.

At fixed mass-squared differences, the dependence on the solar angle
is strictly monotonic. Indeed,
\begin{equation}
\frac{\partial m_1^2}{\partial x}
=
\frac{S}{(1-3x)^2}
>
0.
\label{eq:m1-monotonicity}
\end{equation}
Since the three squared masses have the same derivative with respect
to $x$, it follows explicitly that
\begin{equation}
\frac{\partial\Sigma_\nu}{\partial x}
=
\frac{S}{2(1-3x)^2}
\left(
\frac{1}{m_1}
+
\frac{1}{m_2}
+
\frac{1}{m_3}
\right)
>
0,
\qquad
\lim_{x\to(1/3)^-}\Sigma_\nu
=
+\infty.
\label{eq:mass-sum-monotonicity}
\end{equation}
The proximity of the measured value of $x$ to $1/3$ is therefore
responsible for both the large predicted mass sum and its enhanced
sensitivity to the solar-angle uncertainty.

The frequently used approximation
$\Sigma_\nu\simeq\sqrt{3}\,m_0$ follows from the expansion about the
cubic point. Using the mass expansions in
Eqs.~\eqref{eq:m1-second-order}--\eqref{eq:m3-second-order}, one finds
\begin{equation}
\frac{\Sigma_\nu}{\sqrt{3}\,m_0}
=
1
-
\frac{1}{2}\epsilon_\xi^2
-
\frac{1}{3}\epsilon_\zeta^2
+
\mathcal{O}(\epsilon^3).
\label{eq:mass-sum-cubic-expansion}
\end{equation}
All first-order corrections cancel, explaining why
$\sqrt{3}\,m_0$ provides a numerically accurate estimate. Nevertheless,
Eq.~\eqref{eq:exact-neutrino-mass-sum}, rather than this approximation,
must be used in a precision comparison with cosmology.

For the two input sets introduced in
Eqs.~\eqref{eq:published-JUNO-inputs} and
\eqref{eq:preliminary-JUNO-inputs}, the exact central spectra are
\begin{equation}
\begin{aligned}
\mathcal{D}_{\mathrm{pub}}:\qquad
(m_1,m_2,m_3)
&=
(0.105050,\,0.105407,\,0.116381)\,\mathrm{eV},
\\
\Sigma_\nu
&=
0.326838\,\mathrm{eV},
\\[1mm]
\mathcal{D}_{\mathrm{pre}}:\qquad
(m_1,m_2,m_3)
&=
(0.093761,\,0.094154,\,0.106302)\,\mathrm{eV},
\\
\Sigma_\nu
&=
0.294216\,\mathrm{eV}.
\end{aligned}
\label{eq:absolute-mass-predictions-two-sets}
\end{equation}
The published-solar central value of $x$ lies closer to $1/3$ and
therefore predicts a larger absolute mass scale than the preliminary
value used in Ref.~\cite{Xing2026Cuboid}.

Using the same independent-Gaussian Monte Carlo prescription as in
Sec.~\ref{sec:oscillation}, and retaining only samples in the physical
domain $0<x<1/3$, gives the following medians and central $68\%$
intervals:
\begin{equation}
\begin{aligned}
\mathcal{D}_{\mathrm{pub}}:\qquad
\Sigma_\nu
&=
0.3266^{+0.0806}_{-0.0465}\,\mathrm{eV},
\\
\mathcal{D}_{\mathrm{pre}}:\qquad
\Sigma_\nu
&=
0.2942^{+0.0379}_{-0.0274}\,\mathrm{eV}.
\end{aligned}
\label{eq:mass-sum-MC-intervals}
\end{equation}
The corresponding central $95\%$ intervals are
\begin{equation}
\begin{aligned}
\mathcal{D}_{\mathrm{pub}}:\qquad
0.2495\,\mathrm{eV}
&<
\Sigma_\nu
<
0.5876\,\mathrm{eV},
\\
\mathcal{D}_{\mathrm{pre}}:\qquad
0.2462\,\mathrm{eV}
&<
\Sigma_\nu
<
0.3873\,\mathrm{eV}.
\end{aligned}
\label{eq:mass-sum-MC-95-intervals}
\end{equation}
The pronounced asymmetry, especially for
$\mathcal{D}_{\mathrm{pub}}$, originates from the divergence of
$\Sigma_\nu$ as $x\to(1/3)^-$. These illustrative intervals inherit
the assumptions of Gaussian inputs and vanishing correlations and
should not be interpreted as experimental likelihood intervals for
the alignment hypothesis.

\subsection{Comparison with cosmological limits}

Cosmological constraints on $\Sigma_\nu$ depend on the assumed
cosmological model, the combination of datasets, the treatment of
systematic uncertainties, and the adopted mass prior. Under the
baseline $\Lambda\mathrm{CDM}$ model, the Planck analysis combined with
baryon acoustic oscillation data obtained
\cite{Planck2018Cosmology}
\begin{equation}
\Sigma_\nu
<
0.12\,\mathrm{eV}
\qquad
(95\%~\mathrm{C.L.}).
\label{eq:Planck-mass-sum-bound}
\end{equation}
A more recent DESI analysis combining DR2 baryon acoustic oscillations
with cosmic microwave background data from Planck and the Atacama
Cosmology Telescope reported
\cite{DESI2025Neutrino,Jimenez2026}
\begin{equation}
\Sigma_\nu
<
0.0642\,\mathrm{eV}
\qquad
(95\%~\mathrm{C.L.})
\label{eq:DESI-LambdaCDM-bound}
\end{equation}
within baseline $\Lambda\mathrm{CDM}$ and for three degenerate neutrino
mass eigenstates. The latter result lies close to the
oscillation-imposed normal-ordering floor and is accompanied by a
preference for unphysically small effective neutrino masses in some
statistical treatments. It must therefore be interpreted together with
the reported cosmological-data tensions and prior dependence.

Allowing a time-dependent dark-energy equation of state substantially
relaxes the DESI constraint. For example, in the
$w_0w_a\mathrm{CDM}$ model one obtains
\cite{DESI2025Neutrino}
\begin{equation}
\Sigma_\nu
<
0.163\,\mathrm{eV}
\qquad
(95\%~\mathrm{C.L.})
\label{eq:DESI-dynamical-DE-bound}
\end{equation}
Broader surveys of cosmological assumptions find representative
$2\sigma$ limits distributed around $\Sigma_\nu<0.2\,\mathrm{eV}$
within approximately a factor of three \cite{Capozzi2025}. This spread
demonstrates that cosmological mass limits are not model independent.

Once a cosmological upper limit is treated as a sum-only benchmark,
its algebraic comparison with the exact-alignment prediction is
unambiguous. Transferring a published cosmological likelihood to the
specific nondegenerate spectrum predicted by the alignment is a
separate step and, strictly, requires a cosmological reanalysis. Both
central cuboid predictions in
Eq.~\eqref{eq:absolute-mass-predictions-two-sets} exceed
$0.2\,\mathrm{eV}$, as well as the bounds in
Eqs.~\eqref{eq:Planck-mass-sum-bound}--\eqref{eq:DESI-dynamical-DE-bound}.
Thus, the benchmark bound $\Sigma_\nu<0.2\,\mathrm{eV}$ does not allow
the central exact-alignment spectrum. Limits near the upper end of the
broader model-dependent range, of order $0.6\,\mathrm{eV}$, would allow
it.

Under the simplifying approximation that the relevant cosmological
likelihood depends only on $\Sigma_\nu$, the monotonicity of
$\Sigma_\nu(x)$ allows an upper limit on the mass sum to be translated
into an upper limit on the solar variable $x$. Using the preliminary
central mass-squared differences, one finds
\begin{equation}
\begin{array}{c|cccc}
\Sigma_{\nu,\mathrm{max}}/\mathrm{eV}
&
0.200
&
0.163
&
0.120
&
0.0642
\\
\hline
x_{\mathrm{max}}
&
0.2698
&
0.2393
&
0.1713
&
0.0070
\end{array}.
\label{eq:cosmology-induced-x-bounds}
\end{equation}
For example,
\begin{equation}
\Sigma_\nu<0.2\,\mathrm{eV}
\quad\Longrightarrow\quad
\sin^2\theta_{12}<0.2698.
\label{eq:point-two-eV-solar-bound}
\end{equation}
within the exact alignment. This is already well below both JUNO-based
central values. These translations are algebraic consequences of the
ansatz under the sum-only approximation and at the central
mass-squared differences. They are not combined confidence limits
because the full oscillation and cosmological likelihoods have not
been profiled jointly.

The row based on $0.0642\,\mathrm{eV}$ requires particular caution
because the quoted DESI bound was obtained assuming three degenerate
neutrino masses, whereas the aligned spectrum near this small mass sum
is strongly nondegenerate. This row should therefore be regarded only
as a sum-only illustration, not as a direct likelihood constraint on
the exact-alignment model. More generally, the translation neglects
the dependence of a cosmological likelihood on the distribution of
the total mass among the three eigenstates.

The appropriate conclusion is therefore twofold. The exact alignment
is in strong tension with baseline $\Lambda\mathrm{CDM}$ and with
several moderately extended cosmologies, including the quoted
$w_0w_a\mathrm{CDM}$ analysis. It is not, however, excluded in a
strictly cosmology-independent sense, because sufficiently broad
changes to the cosmological model, the neutrino sector, or the treatment
of systematic uncertainties can weaken the upper limit beyond the
predicted range.

\subsection{Effective mass in beta decay}

The kinematic effective mass measured in beta decay is
\begin{equation}
m_\beta^2
\equiv
\sum_{i=1}^{3}
\lvert U_{ei}\rvert^2m_i^2.
\label{eq:beta-effective-mass-definition}
\end{equation}
Writing $s_{13}^2\equiv\sin^2\theta_{13}$ and
$c_{13}^2\equiv\cos^2\theta_{13}$, the standard parametrization gives
\begin{equation}
m_\beta^2
=
c_{13}^2
\left[
(1-x)m_1^2+xm_2^2
\right]
+
s_{13}^2m_3^2.
\label{eq:beta-effective-mass-expanded}
\end{equation}
Using $m_2^2=m_1^2+\Delta m_{21}^2$ and
$m_3^2=m_1^2+\Delta m_{31}^2$, this simplifies exactly to
\begin{equation}
m_\beta^2
=
m_1^2
+
c_{13}^2x\Delta m_{21}^2
+
s_{13}^2\Delta m_{31}^2.
\label{eq:exact-beta-effective-mass}
\end{equation}
Thus, once $\theta_{13}$ is supplied independently, the alignment also
fixes $m_\beta$.

Taking the representative normal-ordering value
$s_{13}^2=0.0223$ \cite{Capozzi2025}, the two central predictions are
\begin{equation}
\begin{aligned}
\mathcal{D}_{\mathrm{pub}}:\qquad
m_\beta
&=
0.10542\,\mathrm{eV},
\\
\mathcal{D}_{\mathrm{pre}}:\qquad
m_\beta
&=
0.09418\,\mathrm{eV}.
\end{aligned}
\label{eq:beta-mass-two-predictions}
\end{equation}
The KATRIN Collaboration currently reports the direct bound
\cite{KATRIN2025}
\begin{equation}
m_\beta
<
0.45\,\mathrm{eV}
\qquad
(90\%~\mathrm{C.L.}).
\label{eq:KATRIN-current-bound}
\end{equation}
The exact-alignment predictions are therefore fully consistent with
the present direct limit. A future laboratory sensitivity at or below
approximately $0.1\,\mathrm{eV}$ would begin to test the relevant
parameter region without relying on cosmological assumptions.

\subsection{Neutrinoless double-beta decay}

If the three light neutrinos are Majorana particles and the decay is
dominated by the standard light-neutrino-exchange mechanism, the
effective Majorana mass is
\begin{equation}
m_{\beta\beta}
\equiv
\left|
c_{13}^2
\left[
(1-x)m_1
+
xm_2e^{i\alpha_{21}}
\right]
+
s_{13}^2m_3e^{i\alpha_{31}}
\right|,
\label{eq:mbb-exact-definition}
\end{equation}
where $\alpha_{21}$ and $\alpha_{31}$ are the two independent
Majorana-phase combinations.

Define the three positive lengths
\begin{equation}
A
\equiv
c_{13}^2(1-x)m_1,
\qquad
B
\equiv
c_{13}^2xm_2,
\qquad
C
\equiv
s_{13}^2m_3.
\label{eq:mbb-vector-lengths}
\end{equation}
Varying the two phases gives the exact envelope
\begin{equation}
m_{\beta\beta}^{\mathrm{max}}
=
A+B+C,
\label{eq:mbb-maximum}
\end{equation}
and
\begin{equation}
m_{\beta\beta}^{\mathrm{min}}
=
\max
\left[
2\max(A,B,C)-(A+B+C),
\,0
\right].
\label{eq:mbb-minimum-general}
\end{equation}
For both central cuboid spectra, $A$ is the largest contribution, so
the minimum reduces to
\begin{equation}
m_{\beta\beta}^{\mathrm{min}}
=
A-B-C.
\label{eq:mbb-minimum-cuboid}
\end{equation}

Using again $s_{13}^2=0.0223$, the phase-dependent ranges are
\begin{equation}
\begin{aligned}
\mathcal{D}_{\mathrm{pub}}:\qquad
0.03649\,\mathrm{eV}
&\leq
m_{\beta\beta}
\leq
0.10541\,\mathrm{eV},
\\
\mathcal{D}_{\mathrm{pre}}:\qquad
0.03352\,\mathrm{eV}
&\leq
m_{\beta\beta}
\leq
0.09416\,\mathrm{eV}.
\end{aligned}
\label{eq:mbb-cuboid-ranges}
\end{equation}
The nonzero lower limits are important. Although the Majorana phases
can produce substantial destructive interference, they cannot cancel
$m_{\beta\beta}$ arbitrarily close to zero for the exact-alignment
spectrum. The dominant contribution proportional to
$c_{13}^2(1-x)m_1$ is larger than the sum of the other two contributions.

The complete KamLAND-Zen dataset gives
\cite{KamLANDZen2025}
\begin{equation}
T_{1/2}^{0\nu}
>
3.8\times10^{26}\,\mathrm{yr}
\qquad
(90\%~\mathrm{C.L.}),
\label{eq:KamLANDZen-half-life-bound}
\end{equation}
corresponding to the following nuclear-model-dependent range of upper
limits:
\begin{equation}
m_{\beta\beta}
<
(0.028\text{--}0.122)\,\mathrm{eV}.
\label{eq:KamLANDZen-mbb-bound}
\end{equation}
Because the inferred upper limit varies substantially with the adopted
nuclear matrix elements, it does not yet yield a
nuclear-model-independent exclusion of the ranges in
Eq.~\eqref{eq:mbb-cuboid-ranges}. A representative combined analysis
of several isotopes gives $m_{\beta\beta}<0.086\,\mathrm{eV}$ at
$2\sigma$ \cite{Capozzi2025}, which restricts some constructive-phase
configurations but does not exclude the phase-minimized predictions.

Consequently, the exact alignment leads to a sharply correlated set of
nonoscillation observables:
\begin{equation}
\begin{aligned}
\mathcal{D}_{\mathrm{pub}}:\qquad
\left(
\Sigma_\nu,\,
m_\beta,\,
m_{\beta\beta}
\right)
&=
\left(
0.32684,\,
0.10542,\,
0.03649\text{--}0.10541
\right)\,\mathrm{eV},
\\
\mathcal{D}_{\mathrm{pre}}:\qquad
\left(
\Sigma_\nu,\,
m_\beta,\,
m_{\beta\beta}
\right)
&=
\left(
0.29422,\,
0.09418,\,
0.03352\text{--}0.09416
\right)\,\mathrm{eV}.
\end{aligned}
\label{eq:combined-nonoscillation-predictions}
\end{equation}
The cosmological mass sum currently provides the strongest constraint
under standard cosmological assumptions, whereas beta decay supplies
a weaker but cosmology-independent test. If neutrinos are Majorana
particles and standard light-neutrino exchange dominates,
neutrinoless double-beta decay probes an intermediate and experimentally
accessible band. Agreement among these three observables and the
oscillation sum rule would constitute an overconstrained test of the
neutrino-cuboid alignment.

\section{Summary and conclusions}
\label{sec:conclusions}

We have developed the exact phenomenology of the neutrino-cuboid
construction proposed in Ref.~\cite{Xing2026Cuboid}. The cuboid
parametrization itself is a kinematic representation of any positive
three-neutrino mass spectrum. Its predictive physical content arises
only after the geometric angles are identified with the solar and
atmospheric mixing angles through the alignment conditions
$\xi=\theta_{12}$ and $\zeta=\theta_{23}$.

The cubic point corresponds to
$m_1=m_2=m_3=m_0/\sqrt{3}$,
$\xi_*=\arcsin(1/\sqrt{3})$, and $\zeta_*=\pi/4$. The numerical
coincidence of $\xi_*$ and $\zeta_*$ with the solar and atmospheric
angles of tribimaximal mixing does not, by itself, establish a physical
relation between mass degeneracy and flavor mixing. In particular, the
cuboid geometry does not determine $\theta_{13}$, and exact mass
degeneracy makes all vacuum oscillation phases vanish. An additional
mass--mixing alignment hypothesis is therefore essential.

Under the exact alignment, the mass spectrum is determined completely
by $\sin^2\theta_{12}$ and the two measured mass-squared differences.
The usual continuous freedom in the lightest neutrino mass is removed.
The exact relations imply
\begin{equation}
m_0^2
=
\frac{
\Delta m_{21}^2+\Delta m_{31}^2
}{
1-3\sin^2\theta_{12}
},
\qquad
\sin^2\theta_{23}
=
\frac{
x+\eta(1-2x)
}{
(1-x)(1+\eta)
},
\label{eq:conclusion-exact-relations}
\end{equation}
where $x=\sin^2\theta_{12}$ and
$\eta=\Delta m_{21}^2/\Delta m_{31}^2$. These equations are exact and
do not rely on an expansion about the cubic point.

Two qualitative consequences follow immediately. Since the observed
solar angle satisfies $x<1/3$, positivity of $m_0^2$ requires
\begin{equation}
\Delta m_{21}^2+\Delta m_{31}^2>0.
\label{eq:conclusion-ordering-condition}
\end{equation}
In inverted ordering, $\Delta m_{31}^2<0$ and
$\lvert\Delta m_{31}^2\rvert>\Delta m_{21}^2$, so this condition cannot
be satisfied. Inverted ordering is therefore excluded within the exact
ansatz. Moreover,
\begin{equation}
\sin^2\theta_{23}-\frac{1}{2}
=
\frac{
(3x-1)(1-\eta)
}{
2(1-x)(1+\eta)
},
\label{eq:conclusion-octant-relation}
\end{equation}
so $x<1/3$ and $0<\eta<1$ require
$\theta_{23}<\pi/4$. Normal ordering and the lower atmospheric octant
are therefore predictions of the exact alignment rather than
independent assumptions.

For the preliminary JUNO-based input set used in
Ref.~\cite{Xing2026Cuboid}, the exact relations predict
\begin{equation}
\begin{aligned}
\sin^2\theta_{23}^{\mathrm{pred}}
&=
0.439620,
&
\theta_{23}^{\mathrm{pred}}
&=
41.5320^\circ,
\\
(m_1,m_2,m_3)
&=
(0.093761,\,0.094154,\,0.106302)\,\mathrm{eV},
&
\Sigma_\nu
&=
0.294216\,\mathrm{eV}.
\end{aligned}
\label{eq:conclusion-preliminary-predictions}
\end{equation}
The published-solar benchmark, which combines the first published JUNO
solar inputs with the common atmospheric-scale input from the later
preliminary JUNO--Daya Bay combination, instead gives
\begin{equation}
\begin{aligned}
\sin^2\theta_{23}^{\mathrm{pred}}
&=
0.450639,
&
\theta_{23}^{\mathrm{pred}}
&=
42.1672^\circ,
\\
(m_1,m_2,m_3)
&=
(0.105050,\,0.105407,\,0.116381)\,\mathrm{eV},
&
\Sigma_\nu
&=
0.326838\,\mathrm{eV}.
\end{aligned}
\label{eq:conclusion-published-predictions}
\end{equation}
The difference between the two absolute mass scales primarily reflects
the singular sensitivity of the spectrum as
$\sin^2\theta_{12}\to(1/3)^-$.

We have also examined the expansion about the cubic point. Although
the angular deviations are small, the first-order expansion is not
uniformly reliable for every observable. The solar splitting arises
from a cancellation between the leading contributions proportional to
$\epsilon_\xi$ and $\epsilon_\zeta$. For the representative preliminary
solution, the second-order correction to the dimensionless solar
splitting is approximately $49\%$ of its residual first-order value.
When evaluated at the exact angular deviations, the first-order formula
consequently predicts $\eta_{[1]}\simeq0.0560$, almost twice the exact
value $\eta=0.029446$. In contrast, the first-order estimate of $m_0$
differs from the exact result by only approximately $0.6\%$, because
the $\epsilon_\zeta$ dependence cancels from the sum of the two
mass-squared differences. At the same exact deviations, the
second-order treatment reduces the relative error in $\eta$ to
approximately $3.9\%$ and reproduces the exact angular correlation much
more accurately.

The oscillation-data test is presently suggestive but not decisive.
Depending on whether the published-solar or preliminary JUNO-based
benchmark is adopted, and on which representative global determination
of $\theta_{23}$ is used, the central exact-alignment prediction differs
from the global lower-octant best fit by approximately
$0.9\sigma$--$1.9\sigma$ in an illustrative Gaussian comparison. These
numbers are not formal exclusion significances because the relevant
likelihoods are generally asymmetric, the inputs need not be
statistically independent, and the complete covariance information for
the preliminary input is unavailable. A definitive test requires a
joint profile-likelihood analysis imposing the exact map between
$\theta_{12}$, the mass-squared differences, and $\theta_{23}$.

The absolute-mass predictions provide an independent and potentially
stronger test. For the representative value $s_{13}^2=0.0223$, the
preliminary input set predicts
\begin{equation}
m_\beta
=
0.09418\,\mathrm{eV},
\qquad
0.03352\,\mathrm{eV}
\leq
m_{\beta\beta}
\leq
0.09416\,\mathrm{eV},
\label{eq:conclusion-nonoscillation-preliminary}
\end{equation}
where the range for $m_{\beta\beta}$ assumes Majorana neutrinos,
standard light-neutrino exchange, and arbitrary Majorana phases. The
published-solar benchmark instead predicts
\begin{equation}
m_\beta
=
0.10542\,\mathrm{eV},
\qquad
0.03649\,\mathrm{eV}
\leq
m_{\beta\beta}
\leq
0.10541\,\mathrm{eV}.
\label{eq:conclusion-nonoscillation-published}
\end{equation}
Both beta-decay predictions are below the present KATRIN limit. The
Majorana phases can reduce $m_{\beta\beta}$ substantially, but they
cannot produce a complete cancellation because the contribution
proportional to $c_{13}^2(1-x)m_1$ is larger than the sum of the other
two contributions.

Cosmology currently presents the most serious challenge. Both central
predictions for $\Sigma_\nu$ exceed not only the stringent limits
obtained in baseline $\Lambda\mathrm{CDM}$ but also the representative
benchmark $\Sigma_\nu<0.2\,\mathrm{eV}$. Indeed, within the exact
alignment and at the preliminary central mass-squared differences, the
condition $\Sigma_\nu<0.2\,\mathrm{eV}$ would require
$\sin^2\theta_{12}<0.2698$, well below the measured value. The ansatz
is therefore in strong tension with standard cosmological analyses and
with several moderately extended cosmological models. These
comparisons treat the published cosmological limits as sum-only
benchmarks. A rigorous cosmological exclusion of the alignment would
require the relevant likelihoods to be evaluated for its specific
nondegenerate mass spectrum. The conclusion is not cosmology
independent, because the inferred upper limit on $\Sigma_\nu$ can
change substantially when different cosmological models,
neutrino-sector assumptions, datasets, priors, or treatments of
systematic uncertainties are adopted.

The exact neutrino-cuboid alignment is consequently a highly predictive
and readily falsifiable hypothesis. It would be excluded by a
definitive inverted-ordering determination, a definitive upper-octant
value of $\theta_{23}$, a statistically significant violation of the
exact atmospheric-angle sum rule, or a robust absolute-mass constraint
incompatible with its predicted spectrum. Conversely, simultaneous
agreement among the lower-octant oscillation prediction, the
beta-decay mass, and the cosmological mass sum would provide a
nontrivial overconstrained test. If neutrinos are Majorana particles
and standard light-neutrino exchange dominates, the predicted
Majorana-mass band supplies an additional independent test.

If exact alignment is not supported, the residual variables
$R_\xi=\xi_{\mathrm{g}}-\theta_{12}$ and
$R_\zeta=\zeta_{\mathrm{g}}-\theta_{23}$ provide a continuous framework
for quantifying approximate alignment. Such a deformation may remain
phenomenologically useful, but it necessarily sacrifices the exact
ordering, octant, and absolute-mass predictions derived here. The main
value of the exact formulation is therefore not merely its geometric
interpretation, but the sharp network of correlations that allows the
neutrino-cuboid hypothesis to be tested across oscillation experiments,
direct mass searches, neutrinoless double-beta decay, and cosmology.

\appendix

\section{Numerical propagation and consistency checks}
\label{app:numerical}

This appendix specifies the numerical procedure used to propagate the
oscillation-input uncertainties and records several exact consistency
checks. Its purpose is to make the numerical results in
Secs.~\ref{sec:oscillation} and \ref{sec:absolute_mass} reproducible
without introducing additional phenomenological assumptions.

\subsection{Input sampling and physical domain}

We collect the three inputs that determine the exact-alignment
predictions into
\begin{equation}
\mathbf{p}
\equiv
\left(
x,\,
\Delta m_{21}^2,\,
\Delta m_{31}^2
\right).
\label{eq:appendix-input-vector}
\end{equation}
For a general covariance matrix $\mathbf{C}$, the Monte Carlo samples
are drawn according to
\begin{equation}
\mathbf{p}^{(k)}
\sim
\mathcal{N}
\left(
\overline{\mathbf{p}},
\mathbf{C}
\right),
\qquad
k=1,\ldots,N_{\mathrm{MC}}.
\label{eq:appendix-multivariate-sampling}
\end{equation}
In the illustrative analysis presented in the main text, we use
$N_{\mathrm{MC}}=10^6$ and set
\begin{equation}
\mathbf{C}
=
\operatorname{diag}
\left(
\sigma_x^2,\,
\sigma_{21}^2,\,
\sigma_{31}^2
\right),
\label{eq:appendix-diagonal-covariance}
\end{equation}
because a complete covariance matrix covering all three inputs in
either hybrid benchmark, and in particular the parameters of the
preliminary update, is not publicly available. For the asymmetric
atmospheric-scale uncertainty, we use the symmetrized value
$\sigma_{31}=0.026\times10^{-3}\,\mathrm{eV}^2$.

The numerical propagation reported in this work was performed using
Python~3.12.13 and NumPy~2.3.5. For each input set separately, we
initialize \texttt{numpy.random.Generator} with
\texttt{PCG64(20260724)}. Arrays of length $N_{\mathrm{MC}}$ for $x$,
$\Delta m_{21}^2$, and $\Delta m_{31}^2$ are generated in that order.
Samples failing any of the physical-domain conditions below are
discarded by Boolean masking and are not redrawn. Thus,
$N_{\mathrm{MC}}$ denotes the number of initially generated samples,
whereas the number of accepted samples is slightly smaller.

Each sample is required to satisfy
\begin{equation}
0<x^{(k)}<\frac{1}{3},
\qquad
0
<
\left(
\Delta m_{21}^2
\right)^{(k)}
<
\left(
\Delta m_{31}^2
\right)^{(k)},
\label{eq:appendix-physical-domain}
\end{equation}
together with
\begin{equation}
0
<
F\!\left(
x^{(k)},\eta^{(k)}
\right)
<
1,
\qquad
\eta^{(k)}
\equiv
\frac{
\left(
\Delta m_{21}^2
\right)^{(k)}
}{
\left(
\Delta m_{31}^2
\right)^{(k)}
}.
\label{eq:appendix-physical-y-condition}
\end{equation}
The upper bound $x^{(k)}<1/3$ in
Eq.~\eqref{eq:appendix-physical-domain} is essential because the exact
alignment gives $m_0^2<0$ for $x>1/3$ in normal ordering. The positivity
conditions on the sampled mass-squared differences are satisfied with
essentially unit probability for the quoted uncertainties.

If only the Gaussian distribution of $x$ is considered, the
probability of satisfying the physical interval $0<x<1/3$ is
\begin{equation}
P\!\left(
0<x<\frac{1}{3}
\right)
=
\Phi
\left(
\frac{1/3-\overline{x}}{\sigma_x}
\right)
-
\Phi
\left(
-\frac{\overline{x}}{\sigma_x}
\right),
\label{eq:appendix-boundary-acceptance}
\end{equation}
where $\Phi$ denotes the cumulative distribution function of a
standard normal variable. The second term is numerically negligible
for both input sets. The published-solar and preliminary inputs give,
respectively,
\begin{equation}
P_{\mathrm{pub}}
\!\left(
0<x<\frac{1}{3}
\right)
=
0.99723,
\qquad
P_{\mathrm{pre}}
\!\left(
0<x<\frac{1}{3}
\right)
=
0.999998.
\label{eq:appendix-acceptance-probabilities}
\end{equation}
With the generator and sampling prescription specified above, the full
set of physical-domain cuts retains $997\,269$ samples for
$\mathcal{D}_{\mathrm{pub}}$ and $999\,998$ samples for
$\mathcal{D}_{\mathrm{pre}}$.

Conditioning on the physical domain is appropriate when reporting the
distribution of observables under the exact-alignment hypothesis. In a
formal model-comparison or evidence calculation, however, the
likelihood weight outside this domain should not simply be discarded.
It should instead be handled consistently through the constrained
parameter space and the normalization of the likelihood or posterior,
as appropriate to the adopted statistical framework.

\subsection{Transformation to predicted observables}

For every accepted sample, the atmospheric prediction is evaluated
from the exact map
\begin{equation}
y_{\mathrm{pred}}^{(k)}
=
F\!\left(
x^{(k)},\eta^{(k)}
\right)
=
\frac{
x^{(k)}
+
\eta^{(k)}
\left(
1-2x^{(k)}
\right)
}{
\left(
1-x^{(k)}
\right)
\left(
1+\eta^{(k)}
\right)
},
\label{eq:appendix-y-transformation}
\end{equation}
followed by
\begin{equation}
\left(
\theta_{23}^{\mathrm{pred}}
\right)^{(k)}
=
\arcsin\!\sqrt{
y_{\mathrm{pred}}^{(k)}
}.
\label{eq:appendix-theta23-transformation}
\end{equation}
The angle in Eq.~\eqref{eq:appendix-theta23-transformation} is obtained
in radians and is converted to degrees only after the nonlinear
transformation has been performed.

The exact mass scale and squared masses are evaluated as
\begin{equation}
\left(
m_0^{(k)}
\right)^2
=
\frac{
\left(
\Delta m_{21}^2
\right)^{(k)}
+
\left(
\Delta m_{31}^2
\right)^{(k)}
}{
1-3x^{(k)}
},
\label{eq:appendix-m0-transformation}
\end{equation}
and
\begin{equation}
\begin{aligned}
\left(
m_1^{(k)}
\right)^2
&=
x^{(k)}
\left(
m_0^{(k)}
\right)^2,
\\
\left(
m_2^{(k)}
\right)^2
&=
\left(
m_1^{(k)}
\right)^2
+
\left(
\Delta m_{21}^2
\right)^{(k)},
\\
\left(
m_3^{(k)}
\right)^2
&=
\left(
m_1^{(k)}
\right)^2
+
\left(
\Delta m_{31}^2
\right)^{(k)}.
\end{aligned}
\label{eq:appendix-mass-transformation}
\end{equation}
The positive square root is taken for every physical mass. The
cosmological mass sum is then
\begin{equation}
\Sigma_\nu^{(k)}
=
m_1^{(k)}
+
m_2^{(k)}
+
m_3^{(k)}.
\label{eq:appendix-mass-sum-transformation}
\end{equation}
All calculations are performed with unrounded inputs and intermediate
quantities. Rounding is applied only to the final reported results.

For a sampled value of $s_{13}^2$, or for the fixed representative
value used in the main text, the beta-decay observable is computed
from
\begin{equation}
\left(
m_\beta^{(k)}
\right)^2
=
\left(
m_1^{(k)}
\right)^2
+
c_{13}^2x^{(k)}
\left(
\Delta m_{21}^2
\right)^{(k)}
+
s_{13}^2
\left(
\Delta m_{31}^2
\right)^{(k)}.
\label{eq:appendix-beta-transformation}
\end{equation}
The uncertainty in $s_{13}^2$ is neglected in the illustrative
numerical intervals because its contribution is subdominant to that
from the uncertainty in $x$. It can be included straightforwardly by
enlarging $\mathbf{p}$ and its covariance matrix.

\subsection{Quantile-based uncertainty intervals}

For an observable $O$, let $q_p(O)$ denote the empirical quantile such
that a fraction $p$ of the accepted samples lies at or below it. The
quantiles are evaluated using
\texttt{numpy.quantile} with \texttt{method='linear'}. The central
$68\%$ interval is reported as
\begin{equation}
O
=
q_{0.50}(O)
\,{}^{+\left[q_{0.84}(O)-q_{0.50}(O)\right]}
_{-\left[q_{0.50}(O)-q_{0.16}(O)\right]},
\label{eq:appendix-68-quantile-definition}
\end{equation}
and the central $95\%$ interval is
\begin{equation}
q_{0.025}(O)
<
O
<
q_{0.975}(O).
\label{eq:appendix-95-quantile-definition}
\end{equation}
These quantile intervals are preferable to symmetric standard
deviations because the nonlinear transformation to the absolute mass
scale produces strongly asymmetric distributions.

To see the origin of the upper tail, define
\begin{equation}
t
\equiv
1-3x.
\label{eq:appendix-boundary-variable}
\end{equation}
As $t\to0^+$ at fixed positive mass-squared differences,
\begin{equation}
m_i
\propto
t^{-1/2},
\qquad
\Sigma_\nu
\propto
t^{-1/2}.
\label{eq:appendix-boundary-scaling}
\end{equation}
More explicitly, with
$S=\Delta m_{21}^2+\Delta m_{31}^2$ held fixed,
\begin{equation}
\Sigma_\nu
\sim
\sqrt{
\frac{3S}{t}
}
\qquad
\text{as}
\qquad
t\to0^+.
\label{eq:appendix-mass-sum-boundary-asymptotic}
\end{equation}
A Gaussian density for $x$, conditioned on $0<x<1/3$, remains nonzero
arbitrarily close to the upper boundary. Consequently, the second
moment of the induced mass-sum distribution contains the boundary
behavior
\begin{equation}
\left\langle
\Sigma_\nu^2
\right\rangle
\propto
\int_{0}^{\varepsilon}
\frac{\mathrm{d}t}{t},
\label{eq:appendix-divergent-second-moment}
\end{equation}
for any fixed $\varepsilon>0$, and is therefore logarithmically
divergent in the idealized continuously truncated Gaussian model. By
contrast, the first moment remains finite because its boundary
contribution is proportional to
\begin{equation}
\int_{0}^{\varepsilon}
\frac{\mathrm{d}t}{\sqrt{t}}
<
\infty.
\label{eq:appendix-finite-first-moment}
\end{equation}
A finite Monte Carlo sample necessarily produces a finite sample
variance, but that variance is unstable against samples increasingly
close to $x=1/3$. Medians and finite quantiles remain well defined and
therefore provide more robust summaries of the propagated uncertainty.

This behavior does not constitute evidence that the physical neutrino
masses are arbitrarily large. Instead, it shows that oscillation data
modeled by a truncated Gaussian distribution do not, by themselves,
provide a statistically well-behaved upper-mass variance under the
exact-alignment map. The divergence is regulated by independent
absolute-mass information or by a likelihood or prior density that
vanishes sufficiently rapidly as $x\to(1/3)^-$. A non-Gaussian
experimental likelihood does not automatically provide such
regularization if its density remains nonzero at the boundary.

\subsection{Exact numerical consistency checks}

For each accepted sample, the numerical implementation can be checked
using residuals of identities that must hold analytically. In the
following expressions, the Monte Carlo sample label $(k)$ is suppressed
for readability. We define the dimensionless normalization residual
\begin{equation}
r_0
\equiv
\left|
\frac{
m_1^2+m_2^2+m_3^2
}{
m_0^2
}
-1
\right|,
\label{eq:appendix-norm-residual}
\end{equation}
the solar-angle residual
\begin{equation}
r_x
\equiv
\left|
\frac{m_1^2}{m_0^2}
-x
\right|,
\label{eq:appendix-x-residual}
\end{equation}
the mass-splitting residuals
\begin{equation}
r_{21}
\equiv
\left|
\frac{
m_2^2-m_1^2
}{
\Delta m_{21}^2
}
-1
\right|,
\qquad
r_{31}
\equiv
\left|
\frac{
m_3^2-m_1^2
}{
\Delta m_{31}^2
}
-1
\right|,
\label{eq:appendix-splitting-residuals}
\end{equation}
and the atmospheric-angle residual
\begin{equation}
r_y
\equiv
\left|
\frac{
m_2^2
}{
m_2^2+m_3^2
}
-
F(x,\eta)
\right|.
\label{eq:appendix-y-residual}
\end{equation}
In exact arithmetic,
\begin{equation}
r_0
=
r_x
=
r_{21}
=
r_{31}
=
r_y
=
0.
\label{eq:appendix-zero-residuals}
\end{equation}
In double-precision arithmetic, these residuals should remain at the
level expected from floating-point roundoff. They provide direct checks
of the mass normalization, the solar-angle alignment, the two input
splittings, and the atmospheric-angle sum rule.

The relative splitting residuals can be somewhat larger than the
machine precision because they involve subtracting nearly equal
squared masses. Schematically,
\begin{equation}
r_{21}
=
\mathcal{O}
\left(
\epsilon_{\mathrm{mach}}
\frac{m_1^2}{\Delta m_{21}^2}
\right),
\qquad
r_{31}
=
\mathcal{O}
\left(
\epsilon_{\mathrm{mach}}
\frac{m_1^2}{\Delta m_{31}^2}
\right),
\label{eq:appendix-roundoff-amplification}
\end{equation}
where $\epsilon_{\mathrm{mach}}$ is the machine precision. The
amplification is more pronounced for $r_{21}$ because
$\Delta m_{21}^2\ll\Delta m_{31}^2$.

It is therefore numerically preferable to construct $m_2^2$ directly
from $m_1^2+\Delta m_{21}^2$, as in
Eq.~\eqref{eq:appendix-mass-transformation}, rather than reconstructing
the solar splitting from rounded values of $m_1$ and $m_2$. Similarly,
the cancellation-sensitive ratio $\eta$ should be evaluated directly
from the unrounded sampled inputs
$\Delta m_{21}^2/\Delta m_{31}^2$, rather than from reconstructed
differences of squared masses.

\subsection{Majorana-phase envelope}

For completeness, the exact phase envelope of
$m_{\beta\beta}$ can be derived geometrically. With
\begin{equation}
z
\equiv
A
+
Be^{i\alpha_{21}}
+
Ce^{i\alpha_{31}},
\qquad
m_{\beta\beta}
=
\lvert z\rvert,
\label{eq:appendix-mbb-complex-vector}
\end{equation}
define
\begin{equation}
L
\equiv
A+B+C,
\qquad
L_{\mathrm{max}}
\equiv
\max\!\left\{
A,B,C
\right\}.
\label{eq:appendix-mbb-length-definitions}
\end{equation}
The triangle inequality gives
\begin{equation}
m_{\beta\beta}
\leq
L,
\label{eq:appendix-mbb-upper-triangle}
\end{equation}
with equality when all three contributions have the same phase. The
generalized reverse triangle inequality gives
\begin{equation}
m_{\beta\beta}
\geq
\max\!\left\{
2L_{\mathrm{max}}-L,\,
0
\right\}.
\label{eq:appendix-mbb-lower-triangle}
\end{equation}
The two phases $(\alpha_{21},\alpha_{31})$ vary over a connected
domain, and $m_{\beta\beta}$ is a continuous function of them.
Therefore, its image is a connected interval containing both extrema.
It follows that every value between the lower and upper endpoints is
attainable:
\begin{equation}
\max\!\left\{
2L_{\mathrm{max}}-L,\,
0
\right\}
\leq
m_{\beta\beta}
\leq
L.
\label{eq:appendix-mbb-complete-envelope}
\end{equation}
For both representative exact-alignment spectra considered in this
work, $A=L_{\mathrm{max}}$ and $A>B+C$. The lower endpoint consequently
reduces to
\begin{equation}
m_{\beta\beta}^{\mathrm{min}}
=
A-B-C
>
0,
\label{eq:appendix-mbb-positive-minimum}
\end{equation}
establishing that the Majorana phases cannot produce a complete
cancellation for either central spectrum.

\section{Analytic alignment-residual trajectories}
\label{app:residual-trajectories}

The residual formulation introduced in
Eqs.~\eqref{eq:geometric-angle-observables}--\eqref{eq:alignment-origin}
can be developed analytically. It provides a useful geometric
description of departures from exact alignment and an alternative
proof of the mass-ordering obstruction.

Although the geometric parametrization in Sec.~\ref{sec:geometry} was
defined for strictly positive masses, the endpoint
$m_{\mathrm{lightest}}=0$ used below is included as a continuous
boundary limit of the positive-mass domain.

We define the positive quantities
\begin{equation}
\delta
\equiv
\Delta m_{21}^2,
\qquad
\Delta
\equiv
\left|
\Delta m_{31}^2
\right|,
\label{eq:appendix-positive-splittings}
\end{equation}
with $0<\delta<\Delta$, and denote the lightest neutrino mass by
$\mu\equiv m_{\mathrm{lightest}}\geq0$. For fixed oscillation
parameters, varying $\mu$ traces a one-dimensional curve in the
$(R_\xi,R_\zeta)$ plane.

\subsection{Normal-ordering trajectory}

For normal ordering, $\Delta m_{31}^2=\Delta$ and $\mu=m_1$. The three
masses are
\begin{equation}
m_1
=
\mu,
\qquad
m_2
=
\sqrt{
\mu^2+\delta
},
\qquad
m_3
=
\sqrt{
\mu^2+\Delta
}.
\label{eq:appendix-NO-masses}
\end{equation}
The squared norm of the mass vector is
\begin{equation}
m_0^2
=
3\mu^2+\delta+\Delta.
\label{eq:appendix-NO-m0}
\end{equation}
The corresponding geometric angles satisfy
\begin{equation}
\sin^2\xi_{\mathrm{g}}^{\mathrm{NO}}(\mu)
=
\frac{
\mu^2
}{
3\mu^2+\delta+\Delta
},
\label{eq:appendix-NO-xi-squared}
\end{equation}
and
\begin{equation}
\tan^2\zeta_{\mathrm{g}}^{\mathrm{NO}}(\mu)
=
\frac{
\mu^2+\delta
}{
\mu^2+\Delta
},
\qquad
\sin^2\zeta_{\mathrm{g}}^{\mathrm{NO}}(\mu)
=
\frac{
\mu^2+\delta
}{
2\mu^2+\delta+\Delta
}.
\label{eq:appendix-NO-zeta-squared}
\end{equation}
Thus, the normal-ordering residual trajectory is
\begin{equation}
R_\xi^{\mathrm{NO}}(\mu)
=
\arcsin\!\left[
\frac{
\mu
}{
\sqrt{
3\mu^2+\delta+\Delta
}
}
\right]
-
\theta_{12},
\label{eq:appendix-NO-R-xi}
\end{equation}
together with
\begin{equation}
R_\zeta^{\mathrm{NO}}(\mu)
=
\arctan\!\sqrt{
\frac{
\mu^2+\delta
}{
\mu^2+\Delta
}
}
-
\theta_{23}.
\label{eq:appendix-NO-R-zeta}
\end{equation}

Both geometric angles increase monotonically with $\mu$. To see this,
write $u\equiv\mu^2$. Direct differentiation gives
\begin{equation}
\frac{\mathrm{d}}{\mathrm{d}u}
\sin^2\xi_{\mathrm{g}}^{\mathrm{NO}}
=
\frac{
\delta+\Delta
}{
\left(
3u+\delta+\Delta
\right)^2
}
>
0,
\label{eq:appendix-NO-xi-monotonicity}
\end{equation}
and
\begin{equation}
\frac{\mathrm{d}}{\mathrm{d}u}
\sin^2\zeta_{\mathrm{g}}^{\mathrm{NO}}
=
\frac{
\Delta-\delta
}{
\left(
2u+\delta+\Delta
\right)^2
}
>
0.
\label{eq:appendix-NO-zeta-monotonicity}
\end{equation}
Since $u=\mu^2$ is nondecreasing for $\mu\geq0$, these relations
establish the claimed monotonicity.

The endpoints of the trajectory are
\begin{equation}
\begin{aligned}
\mu\to0:\qquad
\xi_{\mathrm{g}}^{\mathrm{NO}}
&\to
0,
&
\zeta_{\mathrm{g}}^{\mathrm{NO}}
&\to
\arctan\!\sqrt{
\frac{\delta}{\Delta}
},
\\
\mu\to\infty:\qquad
\xi_{\mathrm{g}}^{\mathrm{NO}}
&\to
\xi_*,
&
\zeta_{\mathrm{g}}^{\mathrm{NO}}
&\to
\zeta_*.
\end{aligned}
\label{eq:appendix-NO-angle-limits}
\end{equation}
Consequently,
\begin{equation}
\begin{aligned}
\mu\to0:\qquad
\left(
R_\xi^{\mathrm{NO}},
R_\zeta^{\mathrm{NO}}
\right)
&\to
\left(
-\theta_{12},\,
\arctan\!\sqrt{\delta/\Delta}-\theta_{23}
\right),
\\
\mu\to\infty:\qquad
\left(
R_\xi^{\mathrm{NO}},
R_\zeta^{\mathrm{NO}}
\right)
&\to
\left(
\xi_*-\theta_{12},\,
\zeta_*-\theta_{23}
\right).
\end{aligned}
\label{eq:appendix-NO-residual-limits}
\end{equation}
The quasi-degenerate limit therefore approaches the deviations of the
measured angles from the cubic values; it does not generally approach
the exact-alignment origin.

Because $\theta_{12}<\xi_*$, or equivalently $x<1/3$, the equation
$R_\xi^{\mathrm{NO}}(\mu)=0$ has a unique solution. Writing
$x=\sin^2\theta_{12}$, this solution is
\begin{equation}
\mu_\xi^2
=
\frac{
x(\delta+\Delta)
}{
1-3x
}.
\label{eq:appendix-NO-mu-xi}
\end{equation}
This is precisely the exact prediction for $m_1^2$ derived in
Eq.~\eqref{eq:exact-m1}.

Similarly, provided
\begin{equation}
\frac{\delta}{\delta+\Delta}
<
y
<
\frac{1}{2},
\qquad
y
\equiv
\sin^2\theta_{23},
\label{eq:appendix-NO-y-crossing-domain}
\end{equation}
the equation $R_\zeta^{\mathrm{NO}}(\mu)=0$ has the unique solution
\begin{equation}
\mu_\zeta^2
=
\frac{
y\Delta-(1-y)\delta
}{
1-2y
}.
\label{eq:appendix-NO-mu-zeta}
\end{equation}
Within the crossing domains above, exact alignment occurs if and only
if the two coordinate crossings coincide:
\begin{equation}
\mu_\xi^2
=
\mu_\zeta^2.
\label{eq:appendix-simultaneous-crossing}
\end{equation}
Substituting Eqs.~\eqref{eq:appendix-NO-mu-xi} and
\eqref{eq:appendix-NO-mu-zeta} into
Eq.~\eqref{eq:appendix-simultaneous-crossing} gives
\begin{equation}
y
=
\frac{
x+\eta(1-2x)
}{
(1-x)(1+\eta)
},
\qquad
\eta
=
\frac{\delta}{\Delta},
\label{eq:appendix-residual-derivation-sum-rule}
\end{equation}
which reproduces the exact atmospheric-angle sum rule in
Eq.~\eqref{eq:exact-y}. Geometrically, the sum rule is therefore the
condition that the two monotonic residual coordinates vanish at the
same value of the lightest mass.

\subsection{Inverted-ordering obstruction}

For inverted ordering, $\Delta m_{31}^2=-\Delta$ and $\mu=m_3$. The
spectrum is
\begin{equation}
m_1
=
\sqrt{
\mu^2+\Delta
},
\qquad
m_2
=
\sqrt{
\mu^2+\Delta+\delta
},
\qquad
m_3
=
\mu,
\label{eq:appendix-IO-masses}
\end{equation}
and the squared norm of the mass vector is
\begin{equation}
m_0^2
=
3\mu^2+2\Delta+\delta.
\label{eq:appendix-IO-m0}
\end{equation}
The geometric angles obey
\begin{equation}
\sin^2\xi_{\mathrm{g}}^{\mathrm{IO}}(\mu)
=
\frac{
\mu^2+\Delta
}{
3\mu^2+2\Delta+\delta
},
\label{eq:appendix-IO-xi-squared}
\end{equation}
and
\begin{equation}
\tan^2\zeta_{\mathrm{g}}^{\mathrm{IO}}(\mu)
=
\frac{
\mu^2+\Delta+\delta
}{
\mu^2
},
\qquad
\sin^2\zeta_{\mathrm{g}}^{\mathrm{IO}}(\mu)
=
\frac{
\mu^2+\Delta+\delta
}{
2\mu^2+\Delta+\delta
}.
\label{eq:appendix-IO-zeta-squared}
\end{equation}
For $\mu=0$, the first expression in
Eq.~\eqref{eq:appendix-IO-zeta-squared} diverges and is understood
through the limit
$\zeta_{\mathrm{g}}^{\mathrm{IO}}\to\pi/2$.

Subtracting the cubic squared-sine values gives
\begin{equation}
\sin^2\xi_{\mathrm{g}}^{\mathrm{IO}}
-
\frac{1}{3}
=
\frac{
\Delta-\delta
}{
3\left(
3\mu^2+2\Delta+\delta
\right)
}
>
0,
\label{eq:appendix-IO-xi-obstruction}
\end{equation}
and
\begin{equation}
\sin^2\zeta_{\mathrm{g}}^{\mathrm{IO}}
-
\frac{1}{2}
=
\frac{
\Delta+\delta
}{
2\left(
2\mu^2+\Delta+\delta
\right)
}
>
0.
\label{eq:appendix-IO-zeta-obstruction}
\end{equation}
The inequalities follow from $0<\delta<\Delta$. Therefore,
\begin{equation}
\xi_{\mathrm{g}}^{\mathrm{IO}}
>
\xi_*,
\qquad
\zeta_{\mathrm{g}}^{\mathrm{IO}}
>
\zeta_*,
\label{eq:appendix-IO-angles-above-cubic}
\end{equation}
for every finite $\mu$. Since the measured solar angle satisfies
$\theta_{12}<\xi_*$, it follows that
\begin{equation}
R_\xi^{\mathrm{IO}}(\mu)
=
\xi_{\mathrm{g}}^{\mathrm{IO}}(\mu)
-
\theta_{12}
>
\xi_*-\theta_{12}
>
0.
\label{eq:appendix-IO-positive-R-xi}
\end{equation}
The inverted-ordering trajectory can therefore never intersect the
line $R_\xi=0$ and, in particular, cannot reach the exact-alignment
origin. This provides a direct geometric proof of the ordering
obstruction, independent of the sign argument based on $m_0^2$.

If $\theta_{23}<\zeta_*=\pi/4$, the atmospheric residual also satisfies
\begin{equation}
R_\zeta^{\mathrm{IO}}(\mu)
=
\zeta_{\mathrm{g}}^{\mathrm{IO}}(\mu)
-
\theta_{23}
>
\zeta_*-\theta_{23}
>
0.
\label{eq:appendix-IO-positive-R-zeta}
\end{equation}
Thus, for the observed solar angle and a lower-octant atmospheric
angle, the entire inverted-ordering trajectory lies in the
positive-positive quadrant of the residual plane.

\subsection{Restrictions from absolute-mass information}

For normal ordering, the mass sum along the residual trajectory is
\begin{equation}
\Sigma_\nu^{\mathrm{NO}}(\mu)
=
\mu
+
\sqrt{
\mu^2+\delta
}
+
\sqrt{
\mu^2+\Delta
}.
\label{eq:appendix-NO-mass-sum-trajectory}
\end{equation}
It is strictly increasing because
\begin{equation}
\frac{\mathrm{d}\Sigma_\nu^{\mathrm{NO}}}{\mathrm{d}\mu}
=
1
+
\frac{\mu}{\sqrt{\mu^2+\delta}}
+
\frac{\mu}{\sqrt{\mu^2+\Delta}}
>
0.
\label{eq:appendix-NO-mass-sum-monotonicity}
\end{equation}
The minimum mass sum along the trajectory occurs at $\mu=0$ and is
\begin{equation}
\Sigma_{\nu,\mathrm{min}}^{\mathrm{NO}}
=
\sqrt{\delta}
+
\sqrt{\Delta}.
\label{eq:appendix-NO-minimum-mass-sum}
\end{equation}
If a cosmological constraint is represented by a sum-only upper limit
$\Sigma_\nu<\Sigma_\nu^{\mathrm{max}}$ with
$\Sigma_\nu^{\mathrm{max}}>
\Sigma_{\nu,\mathrm{min}}^{\mathrm{NO}}$, it restricts the residual
trajectory to
\begin{equation}
0
\leq
\mu
<
\mu_{\mathrm{max}},
\qquad
\Sigma_\nu^{\mathrm{NO}}
\left(
\mu_{\mathrm{max}}
\right)
=
\Sigma_\nu^{\mathrm{max}}.
\label{eq:appendix-cosmology-trajectory-cut}
\end{equation}
The strict monotonicity of
$\Sigma_\nu^{\mathrm{NO}}(\mu)$ guarantees that
$\mu_{\mathrm{max}}$ is unique. If instead
$\Sigma_\nu^{\mathrm{max}}\leq
\Sigma_{\nu,\mathrm{min}}^{\mathrm{NO}}$, the complete
normal-ordering trajectory is excluded under the same sum-only
interpretation.

The exact-alignment origin survives the cosmological restriction only
if
\begin{equation}
\Sigma_\nu^{\mathrm{NO}}
\left(
\mu_\xi
\right)
<
\Sigma_\nu^{\mathrm{max}},
\label{eq:appendix-origin-cosmology-condition}
\end{equation}
and the angular crossing condition
$\mu_\xi=\mu_\zeta$ is simultaneously satisfied.

For the preliminary central splittings, the illustrative benchmark
$\Sigma_\nu^{\mathrm{max}}=0.2\,\mathrm{eV}$ gives
\begin{equation}
\mu_{\mathrm{max}}
=
0.06044\,\mathrm{eV},
\label{eq:appendix-point-two-mu-max}
\end{equation}
whereas exact alignment requires
\begin{equation}
\mu_\xi
=
m_1
=
0.09376\,\mathrm{eV}.
\label{eq:appendix-preliminary-alignment-mu}
\end{equation}
Thus, under the sum-only interpretation, this benchmark cosmological
bound removes the exact-alignment origin from the allowed segment of
the normal-ordering residual trajectory.

A direct beta-decay bound can be represented similarly. In normal
ordering,
\begin{equation}
m_\beta^2(\mu)
=
\mu^2
+
c_{13}^2x\delta
+
s_{13}^2\Delta.
\label{eq:appendix-beta-mass-trajectory}
\end{equation}
An upper limit $m_\beta<m_\beta^{\mathrm{max}}$ therefore implies
\begin{equation}
\mu^2
<
\left(
m_\beta^{\mathrm{max}}
\right)^2
-
c_{13}^2x\delta
-
s_{13}^2\Delta.
\label{eq:appendix-beta-trajectory-cut}
\end{equation}
If the right-hand side is positive, the allowed trajectory satisfies
\begin{equation}
0
\leq
\mu
<
\sqrt{
\left(
m_\beta^{\mathrm{max}}
\right)^2
-
c_{13}^2x\delta
-
s_{13}^2\Delta
}.
\label{eq:appendix-beta-mu-bound}
\end{equation}
If the right-hand side of
Eq.~\eqref{eq:appendix-beta-trajectory-cut} is nonpositive, the
beta-decay limit excludes the entire normal-ordering trajectory for
the specified mixing parameters and mass-squared differences.
Cosmological and laboratory mass limits can therefore be viewed
geometrically as restrictions on the portion of the residual trajectory
that remains experimentally accessible.

\section*{Data and code availability}

No new experimental data were generated in this work. All numerical
inputs are taken from the published or preliminary results cited in the
main text and are stated explicitly in
Eqs.~\eqref{eq:published-JUNO-inputs},
\eqref{eq:preliminary-JUNO-inputs}, and
\eqref{eq:atmospheric-splitting-input}. The global-fit, direct-mass,
neutrinoless-double-beta-decay, and cosmological constraints used for
comparison are likewise taken from the cited literature.

The exact analytic relations required to reproduce the numerical
predictions are given in the main text. The Monte Carlo sampling
procedure, physical-domain conditions, covariance assumptions, number
and order of generated samples, pseudorandom-number generator, random
seed, rejection prescription, and quantile definitions are specified
in Appendix~\ref{app:numerical}. A complete covariance matrix covering
all three inputs in either hybrid benchmark, and in particular the
parameters of the preliminary JUNO update, is not publicly available.
The illustrative uncertainty propagation therefore assumes independent
Gaussian inputs, as stated explicitly in the analysis.

No specialized or proprietary software was used. The numerical results
reported in this work were generated with Python~3.12.13 and
NumPy~2.3.5, using \texttt{numpy.random.Generator} with
\texttt{PCG64(20260724)}. The calculations require only standard
double-precision floating-point arithmetic and direct implementation of
the equations and sampling prescription provided in this paper. No
separate code repository accompanies this work.

\bibliographystyle{apsrev4-2}
\bibliography{cuboidref}

\end{document}